\documentclass[10pt]{article}

\usepackage[letterpaper]{geometry}
\usepackage{hicss51}
\usepackage{times}
\usepackage[none]{hyphenat}
\usepackage{url}
\usepackage{latexsym}
\usepackage{indentfirst}
\usepackage{graphicx}
\graphicspath{{images/}}
\usepackage{stackengine}
\usepackage{float}
\usepackage{multirow}
\usepackage{multicol}
\usepackage{tabularx}
\usepackage{colortbl}
\usepackage{makecell}
\usepackage[dvipsnames,svgnames]{xcolor}

\title{The impact of differences in facial features between real speakers and 3D face models on synthesized lip motions}

\author{Rabab Algadhy \\
   University of Sheffield\\ United Kingdom \\
  {\underline{ rabab\_elghdi@yahoo.com}} \\\And
  Yoshihiko Gotoh \\
  University of Sheffield\\ United Kingdom \\
  {\underline{y.gotoh@sheffield.ac.uk} }\\\And 
  Steve Maddock \\
   University of Sheffield\\
   United Kingdom  \\
  {\underline{s.maddock@sheffield.ac.uk}} \\}

\date{}

\begin{document}
\maketitle
\begin{abstract}
Lip motion accuracy is important for speech intelligibility, especially for users who are hard of hearing or second language learners. A high level of realism in lip movements is also required for the game and film production industries. 3D morphable models (3DMMs) have been widely used for facial analysis and animation. However, factors that could influence their use in facial animation, such as the differences in facial features between recorded real faces and animated synthetic faces, have not been given adequate attention. This paper investigates the mapping between real speakers and similar and non-similar 3DMMs and the impact on the resulting 3D lip motion. Mouth height and mouth width are used to determine face similarity. The results show that mapping 2D videos of real speakers with low mouth heights to 3D heads that correspond to real speakers with high mouth heights, or vice versa, generates less good 3D lip motion. It is thus important that such a mismatch is considered when using a 2D recording of a real actor’s lip movements to control a 3D synthetic character.
\end{abstract}

\section{Introduction}

The external articulators of the face -- lips, teeth and tongue -- play an important role in facial analysis and animation. They can provide a significant proportion of the visual speech signal \cite{mcgrath1985examination, summerfield1989lips} and improve the intelligibility of speech for hearing impaired people \cite{rogers2006effects, stropahl2017auditory, tyler1997speech} or non-native listeners \cite{hazan2010audiovisual, hazan2005effect} or in adverse listening conditions such as noisy environments \cite{erber1969interaction, middelweerd1987effect,sumby1954visual}. Accurate articulator movement is thus important for visual speech animation and its many application areas. 
There are many approaches to produce visual speech animation \cite{agarwal2021realistic, cohen1993modeling,  liu2020synthesizing, sutskever2014sequence, taylor2012dynamic}. We use a data-driven approach that maps tracked lip motions in 2D videos of a real speaker to corresponding 3D landmarks labelled on a 3D morphable model (3DMM) \cite{blanz1999morphable, egger20193d} built using 3D synthetic head poses. With such a visual speech animation approach, it is important that any mismatches between the real and synthetic faces are considered carefully as a mismatch can reduce the quality of the final animation \cite{taylor2017deep}. Our work focuses on lip motion and how this is affected by similarities and differences in the facial features of the real faces and the 3DMMs. A greater understanding of the impact of any differences could help to provide guidelines for choosing an appropriate real person when animating a non-corresponding synthetic face, i.e. a face that is not the same as the real actor, as might be the case for animating a historical character or a recently-deceased person for a film or game, or even a humanoid character in a film or computer game.

A classification approach for facial features is required. Since we are focusing on lip motion, two mouth features are investigated in this paper: mouth height and width. Based on an analysis of speakers in the Audio-Visual Lombard Grid Speech Corpus \cite{alghamdi2018corpus}, the relevant facial features of each speaker are classified into three categories for each feature: low, middle and high. This is used to investigate the mapping between real speakers and non-corresponding 3DMMs belonging to the same or different classes. 

We believe this is the first paper to provide an extensive study of the impact of a mismatch between the facial features of recorded real faces and synthetic 3D faces on the resulting 3D lip motions. We show that when 2D videos of real speakers who have low mouth height are mapped to 3D synthetic heads that correspond to real speakers who have high mouth height, or vice versa, that the resulting animation is adversely affected. Both quantitative and qualitative evaluations are included in the results.

The remainder of this paper is structured as follows. Section \ref{related_work} gives an overview of related work. Section \ref{methodolgy} presents metrical analysis of facial features for the speakers in the audio-visual Lombard Grid speech corpus and summarises the process of mapping 2D video data to a 3DMM. Sections \ref{subsec:Objective_Evaluation_Ex6} and \ref{subsec:Subjective_Evaluation_Ex6} address quantitative and qualitative evaluation, respectively. Section \ref{sec:conclusion} presents conclusions.

\section{Related work}
\label{related_work}
Generally, based on the input data to a synthetic speech generator, audiovisual speech synthesis approaches can be classified into two main categories: viseme driven approaches and data driven approaches. Viseme driven approaches involve segmenting audio speech signals into phonemes which are then classified into visual units called visemes. Viseme parameters are then interpolated with co-articulation rules incorporated \cite{cohen1993modeling,edwards2016jali, pelachaud1996generating} or using dominance functions \cite{cohen1993modeling, massaro201212} or deep learning approaches \cite{chung2017you, chai2022speech, choudhury2023review, fan2022faceformer, karras2017audio, vougioukas2020realistic, zhu2021deep}. The previous approaches are based on determining the weight of the target phoneme against the neighboring segments and their influence on the corresponding control parameters \cite{cohen1993modeling}. The rule-based models \cite{pelachaud1996generating, beskow1995rule, pelachaud1991communication} take into account only visemes that have an impact on the neighbouring ones (backward and forward coarticulations). However, these models typically fail to fully take co-articulation effects into account, thus leading to unrealistic lip motions. For this reason, data-driven approaches are favored more as they are based on animating faces according to captured data from real speakers, which guarantee coarticulation effects are considered.

Data driven approaches involve capturing facial motion data occurring in actual speech, which is then linked with a 3D face model either based on phonetic information rules (sample-based approaches) \cite{taylor2012dynamic, bregler1997video,cao2004real, cosatto1998sample,  graf2000face, kawai2015automatic, kshirsagar2003visyllable,  xing2023codetalker}, or by using statistical models to control the facial motion that is learned from the training data (learning-based approaches) \cite{karras2017audio, garrido2016corrective,  pham2018end, yu2016real, yu2019deep}. 

These approaches require performance tracking for real faces and then the reconstruction of corresponding 3D faces using the tracked data to produce smooth mouth animations with co-articulation effects. Thus, the realism of the resulting speech animation is highly dependent on how well the 3D face model is reconstructed.  There are two main traditional approaches used for reconstructing 3D face models. The first is based on capturing performance of real faces using RGB or RGB-D cameras and then reconstructing the 3D face model using the captured data. The other approach involves scanning and then blending 3D real faces \cite{lewis2014practice}.
Comprehensive research has been conducted on reconstructing 3D face models from optical sensor measurements of a subject's performance. Face performance can be captured from the subject and represented in a 3D domain either based on illumination data only \cite{ghosh2011multiview, ma2007rapid} or with the aid of markers \cite{bickel2007multi, 10061572, huang2004hierarchical, wood20223d}. These techniques involve capturing the face using a camera and then reconstructing the face geometry either via triangulation or colour and depth values based on the type of camera used. In the past, calibrated dense camera arrays with complex indoor lighting setups, which are expensive to set up and operate, were used \cite{huang2004hierarchical, beeler2011high, fyffe2014driving}. Recently, low-cost devices such as monocular RGB and RGB-D cameras have been used for offline and online monocular face reconstruction and tracking \cite{garrido2016reconstruction, thies2015real, thies2016face2face}. However, the quality of these techniques is affected by lighting conditions that may produce undesired pixels with noisy depth values. This makes capturing faces more complicated. In addition, these techniques are highly challenging since they are based on forming the image by convolving multiple physical dimensions in a single colour measurement. Therefore, current state-of-the-art approaches employ face models and statistical analysis of the distribution of 3D facial shapes.

A large body of research has been conducted on modelling the structure and expression of faces based on low-dimensional subspaces. In some works, a blendshape expression model based on a set of 3D face models, each representing a particular expression, has been used. Facial animation can be achieved by morphing between the neutral face and a specific expression, or by morphing between various expressions. Several studies used delta variations to linearly add each expression to the neutral face \cite{garrido2016corrective, thies2016face2face, bouaziz2013online,  li2013realtime,  weise2011realtime}. Blendshape models can be constructed using multi-monocular cameras in general surroundings \cite{garrido2016corrective}, monocular RGB cameras \cite{thies2016face2face} or RGB-D sensor devices \cite{bouaziz2013online}. Although these methods can achieve globally pleasant results with regards to the static realism of the face rendering, all of these approaches require professional camera setups for gathering data to train the blendshape model, and the lip shapes that can be acquired during speech are still not fully included.

There is another body of research based on the most commonly used prior, presented by Blanz and Vetter  \cite{blanz1999morphable}, which constructs a 3D face model by learning from a low-dimensional face subspace created from high-resolution laser scans of real faces with neutral expressions (3D morphable model (3DMM)). Geometry and the illumination-corrected textures of the faces are included in such models. These methods applied this reconstruction for 2D face recognition, pose normalisation and illumination \cite{huber2016multiresolution, koppen2018gaussian}, face reanimation  \cite{blanz2003reanimating} and facial expression tracking \cite{amberg2008expression} in 2D images, but they were rarely extended to track lip motions during speech  \cite {cudeiro2019capture, musti20143d} due to the expensive devices that are required for gathering the data from real speakers, in addition to the complexity of techniques for preprocessing the gathered data. These methods lack person-specific facial characteristics. In addition, they ignore anatomical and physical plausibility in the reconstructions. Some work has been conducted to personalize the model, either by increasing the number of 3D laser scans to model the skin reflectance, sex and age variations \cite{huber2016multiresolution, koppen2018gaussian, booth2018large, booth20163d}, or by fitting person-specific shape correctives to the generic face models \cite{bouaziz2013online, li2013realtime}, where performance of the resulting animation is mainly limited to noise levels and resolution of the input device. However, when using 3DMMs for producing facial animation, the impact of facial feature differences between real faces and 3D synthetic faces on the resulting 3D animation does not appear to have been considered. There is a gap in the literature that our work aims to address.

\section{Methodology}
\label{methodolgy}
Our approach tracks lip motions in 2D videos of real speakers and maps tracked points to equivalent landmark points labelled on a 3D morphable model (3DMM), built using 3D synthetic head poses, to produce animation. The 2D videos from the Audio-Visual Lombard Grid Speech Corpus \cite{alghamdi2018corpus} are used as input. Section \ref{sec:Facial_features_classification} describes this input data, how it is landmarked and how this is used to classify the faces. Section \ref{3DMM} describes the approach used to construct a 3DMM for each real face used from the dataset. Section \ref{subsec:mapping_process} describes the mapping process from 2D video to 3DMM.

\subsection{Input data and classification of facial features}
\label{sec:Facial_features_classification}
To analyse the facial features of the Audio-Visual Lombard Grid Speech Corpus’s speakers \cite{alghamdi2018corpus}, a method presented by Roelfose et al. \cite{roelofse2008photo} was used. They used morphometrical methods to classify the facial features of South African males in photos to investigate common and rare features in this community. The method is based on both measurements and morphology of the face, which provides a reliable procedure for facial features classification.

The Audio-Visual Lombard Grid Speech Corpus consists of both front- and side-view video of 54 speakers (30 female and 24 male) uttering sentences from the GRID corpus \cite{cooke2006audio} in both plain and Lombard conditions. Only front-view videos of plain sentences are used from the corpus for mapping between 2D videos of a real speaker and a 3DMM, while both front- and side-view videos are used for creating the 3DMMs.

A pool of 27 speakers (12 male and 15 female) of the Audiovisual Lombard Grid Speech corpus was selected to validate the performance of the 3D head models. This pool contains videos of real speakers whose faces are not obscured by glasses or facial hair, heads are not tilted downward, and bottom chin points are visible. The purpose of this step was to facilitate the facial landmark annotation process.

\begin{figure}[htb]
\begin{center}
    {\includegraphics[width=0.45\columnwidth, keepaspectratio]{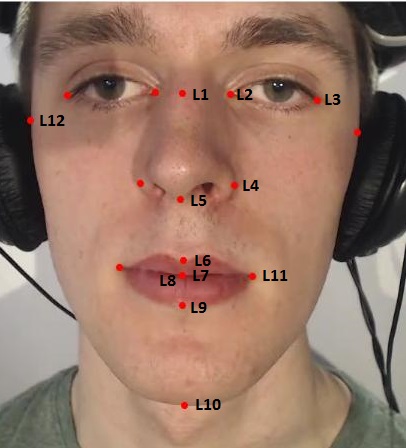}}{}
    {\includegraphics[width=0.45\columnwidth, keepaspectratio]{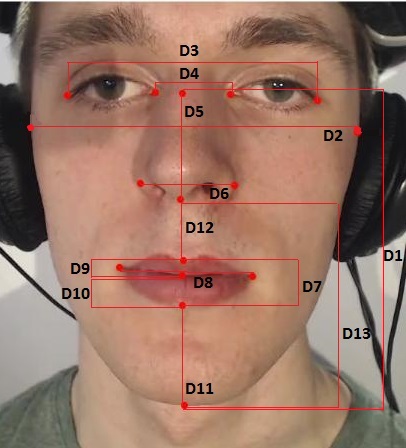}}{}
    \caption{Face landmarks (left) and measurements used for each video frame (right).}
    \label{Fig:Landmarks_Classification}
\end{center}
\end{figure}

Because lip shapes are affected by facial movements such as smiling and crying, the lips must be assessed when the subject has a neutral face shape \cite{carey2009elements}. Thus, videos of the corpus were investigated for each speaker (27 speakers) to select the appropriate frame. The selected frames were processed using Faceware Analyzer software to obtain x and y coordinates for each landmark. Figure \ref{Fig:Landmarks_Classification} (left) shows the landmarks (Lx) that were utilised to take the facial measurements. Figure  \ref{Fig:Landmarks_Classification} (right) shows the 13 measurements (Dx) that were taken from each frame using the Euclidean distance between the predetermined facial landmarks. These distances were used to create indices, irrespective of camera distance. Table \ref{tab:Metrical_features} shows how the twelve indices were computed using the calculated measurements.

\begin{table}[h!]
\begin{center}    

\begin{tabular}{|l|c|}
\hline
\multicolumn{1}{|c|}{Index} & Calculation   \\ \hline
I1-Facial& (100 * (D1/D2)) \\ \hline
I2-Intercanthal &(100 * (D4/D3))  \\ \hline
I3-Nasal& (100 * (D6/D5)) \\ \hline
I4-Nasofacial&  (100 * (D5/D1) \\ \hline
I5-Nose-face width &(100 * (D6/D2))\\ \hline
I6-Lip area&(100 * (D7/D8)) \\ \hline
I7-Vertical mouth height&(100 * (D7/D1))\\ \hline
I8-Upper lip thickness&(100 * (D9/D1)) \\ \hline
I9-Lower lip thickness&(100 * (D10/D1)) \\ \hline
I10-Mouth width&(100 * (D8/D3)) \\ \hline
I11-Chin size&(100 * (D11/D1))\\ \hline
I12-Nose-upper-lips& (100 * (D12/D13)) \\ \hline
\end{tabular}
\caption{Metrical features (indices) for the speakers in the audio-visual Lombard grid speech corpus.}
\label{tab:Metrical_features}
\end{center}
\end{table}

\begin{figure*}[htb]
\begin{center}
    {\includegraphics[trim = 03mm 10mm 2mm 5mm, clip,width=6.60in,keepaspectratio]{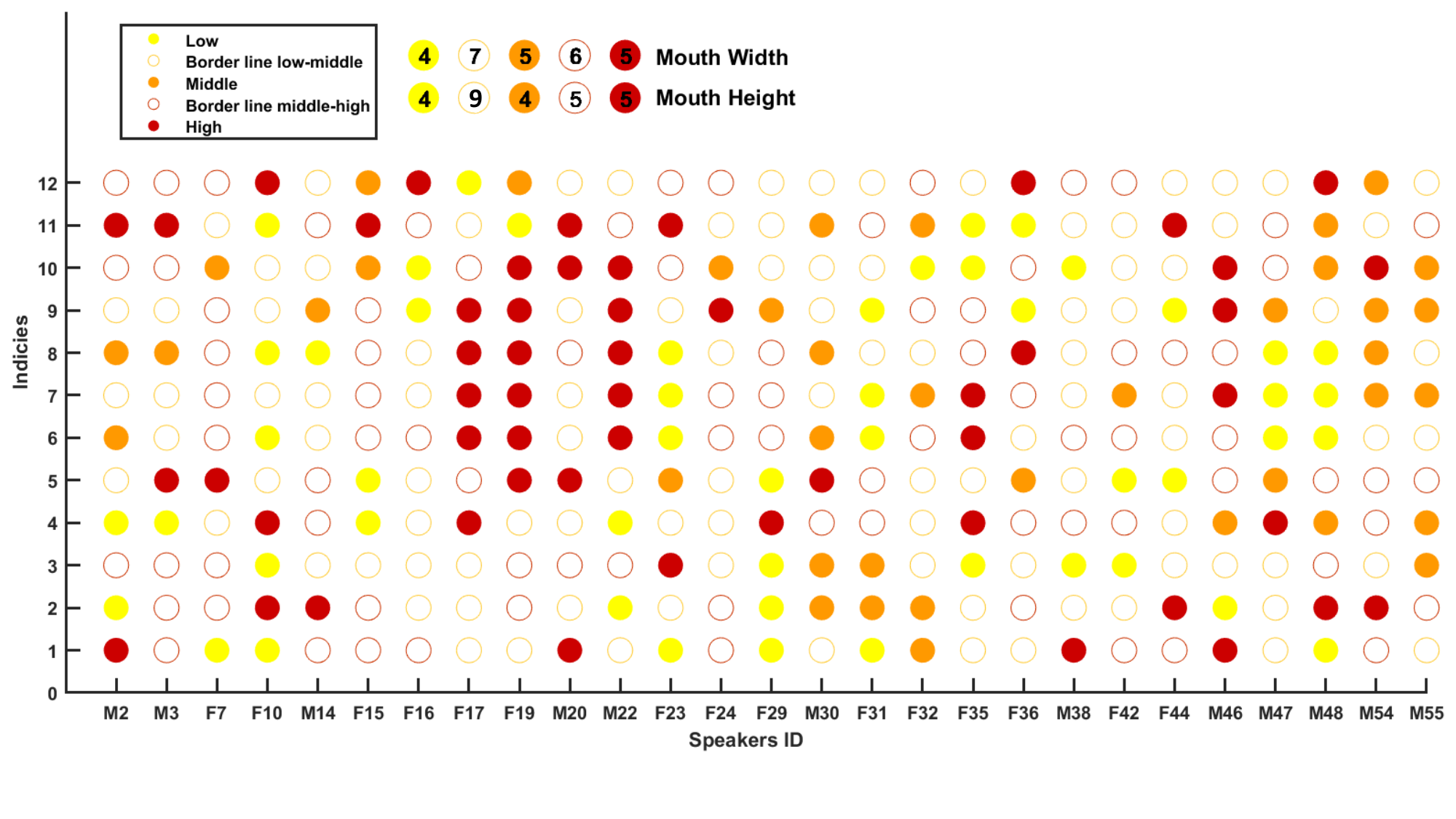}}{}
    \caption{Classification of indices for each speaker of the audio-visual Lombard grid speech corpus (80\%  confidence level for each class of each index). The x axis shows the speaker's ID (where M refers to male speaker and F refers to female speaker) and the y axis shows the indices' number. Number of speakers in each class of mouth height and mouth width indices are shown in the relevant circles at the top of the figure.}
    \label{fig:Method B_Classi80}
\end{center}
\end{figure*}

The ranges of each index were used to categorise the features into three different morphological classes, low, middle and high, using the interquartile range (IQR) and outliers (calculated using Q1-1.5*IQR and Q3+1.5*IQR, where Q1 and Q3 are the lower and upper quartile values, respectively). The classification of the twelve indices for each speaker in the corpus was calculated with 80\% of the confidence intervals for each class of each index. Figure \ref{fig:Method B_Classi80} shows the classification of the twelve indices for the selected 27 real speakers from the audio-visual Lombard grid corpus, where the low, middle and high classes are represented in yellow, orange and red filled circles respectively. The number of speakers in each class of mouth height index and mouth width index are shown in the relevant circles. 

\subsection{3D morphable model}
\label{3DMM}

A 3DMM is created for each real speaker used from the input data set. The real speakers used were the selection classified under one of the three classes of index 7 (mouth height) and index 10 (mouth width), as presented in Figure \ref{fig:Method B_Classi80} that shows the three classes (low, middle and high, represented in filled yellow, orange and red circles, respectively). The 3DMM is constructed using synthetic faces. The commercial software FaceGen is used to create 161 head poses, consisting of a neutral pose and 10 intensity variations of 16 visemes, where each head mesh has vertex correspondence. The tongue and teeth meshes are excluded. The initial neutral head pose can be generated using photo images of a real person by placing facial landmarks either on a front-view only photo or front- and side-view photographs. In this paper, the front- and side-view photographs of each real speaker are used. Then the software is used to deform the face into the required range of poses, where 161 poses are used to train the 3DMM as explained in our previous work \cite{algadhy20193d}.  

Principal Component Analysis (PCA) is applied to the vertices of the 161 Facegen head poses to generate a 3DMM. It is not applied to the texture coordinates since all the Facegen heads have the same texture. The geometry of the head is represented by a shape vector $S=(X_1,Y_1,Z_1,\ldots,X_n,Y_n,Z_n)^\top$, containing the $X$, $Y$, $Z$ coordinates of 5850 vertices. The 3DMM consists of a PCA model of the shape, which is represented as:
\begin{equation}
  M := \{\overline{F},\sigma,V\}
\end{equation}
where $\overline{F}\in R^{3N}$ is the mean vector of the example meshes (mean head) with $N$ being the number of mesh vertices. $\sigma\in R^{n-1}$ denotes the standard deviation, where $n$ is the number of Facegen heads used to build the 3DMM, and $V = [v_{1},.....,v_{n-1}]\in R^{3N\times n-1}$ is a set of principal components in the model.

A new head can be generated as follows:
\begin{equation}
  S = \overline{F}+ \sum_{i=1}^K \alpha_i \sigma_i v_i
\end{equation}
where $K \leq n-1$ is the number of principal components and $\alpha_i \in R^K$ is the shape coefficient \cite{huber2016multiresolution}.

\subsection{Mapping process}
\label{subsec:mapping_process}

The mapping between the 2D frames of a video and a 3DMM is achieved using the camera matrix method presented by Huber et al \cite{huber2016multiresolution}.  In order to track the facial features of a speaker in video, the random cascaded-regression copse (R-CR-C) approach presented by Feng et al \cite{feng2015random} is applied to generate a learned landmark detection model using the Ibug-Helen test set \cite{sagonas2016300}.

Given 51 2D landmarks (of facial features used to animate the 3D mesh) and the corresponding 3D landmarks, a pose of the face is estimated using the Gold Standard Algorithm (more details in \cite{huber2016multiresolution}). It computes the camera matrix that is used to reconstruct the 3D shape. The most likely vector of PCA shape coefficients $\alpha$ is found by minimising the following cost function:
\begin{equation}
  E = \sum\limits_{i=1}^{3N}\frac{(y_{m2D,i}-y_{i})^2}{2\sigma_{2D}^2} + \Vert\alpha\Vert_2^2  
\end{equation}
where $N$ is the number of landmarks, $y$ is the 2D landmarks represented in homogeneous coordinates, $\sigma_{2D}^2$ is an ad hoc variance of these landmarks, and $y_{m2D,i}$ is the projected 3D landmarks to a 2D plane using the camera matrix \cite{huber2016multiresolution}. 

\section{Quantitative evaluation}
\label{subsec:Objective_Evaluation_Ex6}

\begin{figure}[h]
\begin{center}
\stackunder[5pt]{\includegraphics[trim = 21mm 00mm 25mm 00mm, clip,width=3.50in,keepaspectratio]{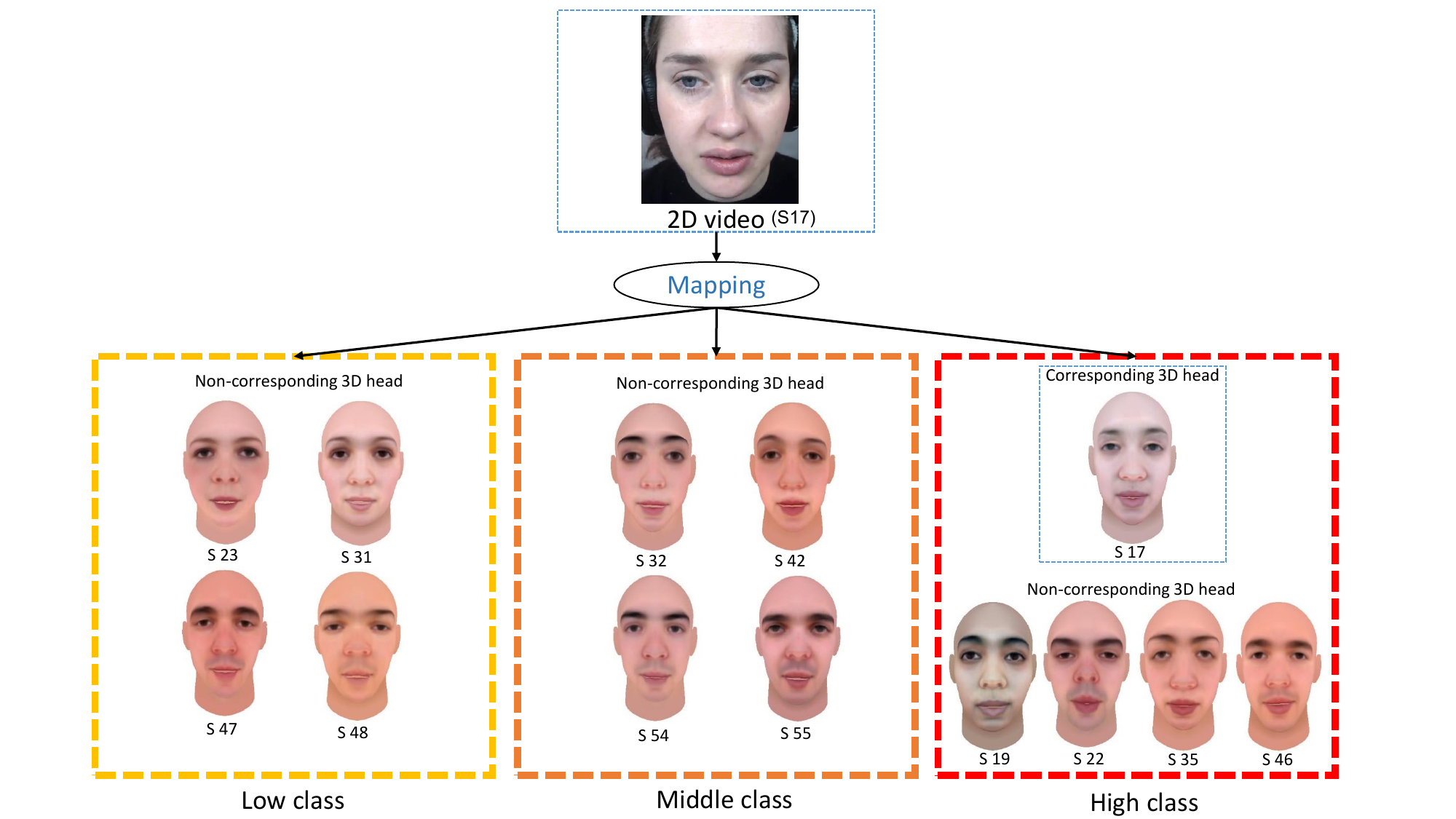}}{}
\caption{An example of the mapping process between 2D video frames of a real speaker (ID: S17) who classified under the high class of index 7,   the corresponding 3D head, and the non-corresponding 3D heads.}
\label{Fig:MappingProcess_S17}
\end{center}
\end{figure}

For each speaker from the selected pool (27 speakers), four plain sentences from the front-view video files were chosen to be mapped to the corresponding 3D head and the non-corresponding 3D heads. The chosen sentences contain different words (e.g different verbs (bin, lay, place and set) and letters (a, b, etc.)), in order to contain the maximum number of English phonemes. For example, referring to index 7 in Figure \ref{fig:Method B_Classi80}, a real speaker (ID: S17) was classified into the high class. 2D videos of this speaker were mapped to the corresponding 3D head and the non-corresponding 3D heads of other speakers in the same class (high class) (i.e. speaker IDs: S19, S22, S35 and S46), middle class (i.e. speaker IDs: S32, S42, S54 and S55) and low class (i.e. speaker IDs: S23, S31, S47 and S48) (see Figure \ref{Fig:MappingProcess_S17}).  Next, 2D videos of the resulting 3D lip motions of each head were compared with the original ground-truth 2D videos. For comparison, Faceware Analyzer software was used to track the facial features in the ground-truth 2D video and the front-view (2D) of the 3D animation. This software can process video files as a batch, which speeds the evaluation process. Two geometric articulatory measurements were calculated from the extracted facial features.  The first was a width measurement defined by the horizontal distance between the right and left inner corners of the lips. The second was a height measurement defined by the distance between the top and the bottom middle of the inner mouth contour.

In order to correct the distance between the camera and the real speaker or the talking 3D head, all the landmarks were normalised using the Euclidean distance between the midpoint of the inner corners of the eyes and the nose tip, since it is assumed these are not affected by the articulations \cite{celiktutan2013comparative}.  All visual articulatory features for the real speakers and their corresponding 3D heads were normalised by their corresponding maximum and minimum mouth measurements in the videos. Given the height and width values for each frame of animation,  for both the real video for a speaker and the resulting 3D animation, the root mean square error (RMSE) over a sentence was used to evaluate the effectiveness of each 3DMM.

\subsection{Mouth height (index 7)}
\label{subsub:index7}
Table \ref{tab:Low_In7} shows the RMSE results averaged over four sentences for the width and height of the mouth aperture of real speakers in the low class of index 7 and their corresponding, non-corresponding low, non-corresponding middle and non-corresponding high 3D heads. From this table it is clear that the RMSE results varied when 2D videos of real speakers were mapped to non-corresponding low 3D heads or non-corresponding middle 3D heads. When the 2D videos were mapped to the non-corresponding high 3D heads, the corresponding 3D heads gave the lowest RMSE scores for height for all speakers and for width for two out of four speakers who were classified in the low to middle class (ID: S31) and the middle class (IDs: S48) of index 10 (mouth width). The corresponding 3D heads of the real speakers (IDs: S23 and S47) failed to give the lowest score for width because their corresponding real speakers were classified under the middle to high class of index 10, which is very close to most of the non-corresponding high 3D heads.

For the corresponding low 3D head of each real speaker and the non-corresponding low 3D heads, a t-test suggested no significant difference in RMSE results for width and height. Additionally, no significant difference was found between the corresponding low 3D heads and the non-corresponding middle 3D heads for all speakers for height and three out of four speakers for width. The significant difference in width given by the corresponding low 3D head of a real speaker (ID: S31) was due to its low mouth width. A significant difference was found between three out of four of the corresponding low 3D heads and the non-corresponding high 3D heads for height. The corresponding low 3D head of a real speaker (ID: S48) suggested no significant differences. This may be due to the large distance between the nose tip and the upper lip (index 12), which reduces the height of the mouth aperture. For width, three out of four of the corresponding 3D heads showed a significant difference. The corresponding low 3D head of a real speaker(ID: S23) suggested no significant difference; this may be because it was classified in the middle to high class of indices 10 and 12.

\begin{table*}[]
\begin{center}
\scalebox{0.80}{
\begin{tabular}{cccccccccccllcc}
\hline
                            & \multicolumn{2}{c}{}                                                                                        & \multicolumn{10}{c}{\begin{tabular}[c]{@{}c@{}}Non-corresponding \\ 3D head (low)\end{tabular}}                                                                                                                                                                                                      & \multicolumn{2}{c}{}                                                                             \\ \cline{4-13}
                            & \multicolumn{2}{c}{\multirow{-2}{*}{\begin{tabular}[c]{@{}c@{}}Corresponding 3D \\ head (low)\end{tabular}}} & \multicolumn{2}{c}{S23}                                                   & \multicolumn{2}{c}{S31}                            & \multicolumn{2}{c}{S47}                            & \multicolumn{2}{c}{S48}                            & \multicolumn{2}{l}{\cellcolor[HTML]{C0C0C0}}          & \multicolumn{2}{c}{\multirow{-2}{*}{\begin{tabular}[c]{@{}c@{}}T-test\\ (P value)\end{tabular}}} \\ \cline{2-15} 
\multirow{-3}{*}{2D video} & W                                                    & H                                                    & W                                               & H                        & W                        & H                        & W                        & H                        & W                        & H                        & \cellcolor[HTML]{C0C0C0}  & \cellcolor[HTML]{C0C0C0}  & W                                               & H                                              \\ \hline
S23                        & 0.270                                                & 0.131                                                & \cellcolor[HTML]{C0C0C0}{\color[HTML]{C0C0C0} } & \cellcolor[HTML]{C0C0C0} & 0.232                    & \textbf{0.114}           & 0.253                    & 0.137                    & \textbf{0.188}           & 0.145                    & \cellcolor[HTML]{C0C0C0}  & \cellcolor[HTML]{C0C0C0}  & 0.1399                                          & 0.9195                                         \\ \hline
S31                        & 0.193                                                & 0.163                                                & 0.208                                           & 0.201                    & \cellcolor[HTML]{C0C0C0} & \cellcolor[HTML]{C0C0C0} & \textbf{0.189}           & 0.229                    & 0.219                    & \textbf{0.125}           & \cellcolor[HTML]{C0C0C0}  & \cellcolor[HTML]{C0C0C0}  & 0.2946                                          & 0.5523                                         \\ \hline
S47                        & 0.218                                                & 0.099                                                & 0.212                                           & \textbf{0.081}           & 0.227                    & 0.149                    & \cellcolor[HTML]{C0C0C0} & \cellcolor[HTML]{C0C0C0} & \textbf{0.140}           & 0.082                    & \cellcolor[HTML]{C0C0C0}  & \cellcolor[HTML]{C0C0C0}  & 0.4501                                          & 0.8448                                         \\ \hline
S48                        & 0.149                                                & \textbf{0.071}                                       & \textbf{0.131}                                  & 0.101                    & 0.157                    & 0.083                    & 0.188                    & 0.073                    & \cellcolor[HTML]{C0C0C0} & \cellcolor[HTML]{C0C0C0} & \cellcolor[HTML]{C0C0C0}  & \cellcolor[HTML]{C0C0C0}  & 0.6168                                          & 0.2153                                         \\ \hline
\multicolumn{1}{l}{}        & \multicolumn{1}{l}{}                                 & \multicolumn{1}{l}{}                                 & \multicolumn{10}{c}{\begin{tabular}[c]{@{}c@{}}Non-corresponding \\ 3D head (middle)\end{tabular}}                                                                                                                                                                                                   & \multicolumn{1}{l}{}                            & \multicolumn{1}{l}{}                           \\ \cline{4-13}
\multicolumn{1}{l}{}        & \multicolumn{1}{l}{}                                 & \multicolumn{1}{l}{}                                 & \multicolumn{2}{c}{S32}                                                   & \multicolumn{2}{c}{S42}                            & \multicolumn{2}{c}{S54}                            & \multicolumn{2}{c}{S55}                            & \multicolumn{2}{l}{\cellcolor[HTML]{C0C0C0}}          & \multicolumn{1}{l}{}                            & \multicolumn{1}{l}{}                           \\ \hline
\multicolumn{1}{l}{}        & W                                                    & H                                                    & W                                               & H                        & W                        & H                        & W                        & H                        & W                        & H                        & \cellcolor[HTML]{C0C0C0}  & \cellcolor[HTML]{C0C0C0}  & W                                               & H                                              \\ \hline
S23                        & 0.270                                                & \textbf{0.131}                                       & \textbf{0.241}                                  & 0.135                    & 0.256                    & 0.168                    & 0.265                    & 0.185                    & 0.277                    & 0.152                    & \cellcolor[HTML]{C0C0C0}  & \cellcolor[HTML]{C0C0C0}  & 0.2696                                          & 0.0734                                         \\ \hline
S31                        & \textbf{0.193}                                       & 0.163                                                & 0.229                                           & \textbf{0.159}           & 0.281                    & 0.206                    & 0.245                    & 0.218                    & 0.248                    & 0.213                    & \cellcolor[HTML]{C0C0C0}  & \cellcolor[HTML]{C0C0C0}  & 0.0132                                          & 0.0767                                         \\ \hline
S47                        & 0.218                                                & 0.099                                                & 0.250                                           & \textbf{0.091}           & 0.234                    & 0.131                    & \textbf{0.170}           & 0.103                    & 0.173                    & 0.098                    & \cellcolor[HTML]{C0C0C0}  & \cellcolor[HTML]{C0C0C0}  & 0.6233                                          & 0.4975                                         \\ \hline
S48                        & 0.149                                                & \textbf{0.071}                                       & 0.157                                           & \textbf{0.071}           & 0.187                    & 0.078                    & \textbf{0.139}           & 0.104                    & 0.166                    & 0.072                    & \cellcolor[HTML]{C0C0C0}  & \cellcolor[HTML]{C0C0C0}  & 0.2762                                          & 0.2772                                         \\ \hline
\multicolumn{1}{l}{}        & \multicolumn{1}{l}{}                                 & \multicolumn{1}{l}{}                                 & \multicolumn{10}{c}{\begin{tabular}[c]{@{}c@{}}Non-corresponding\\ 3D head (high)\end{tabular}}                                                                                                                                                                                                      & \multicolumn{1}{l}{}                            & \multicolumn{1}{l}{}                           \\ \cline{4-13} 
\multicolumn{1}{l}{}        & \multicolumn{1}{l}{}                                 & \multicolumn{1}{l}{}                                 & \multicolumn{2}{c}{S17}                                                   & \multicolumn{2}{c}{S19}                            & \multicolumn{2}{c}{S22}                            & \multicolumn{2}{c}{S35}                            & \multicolumn{2}{c}{S46}                              & \multicolumn{1}{l}{}                            & \multicolumn{1}{l}{}                           \\ \hline
                            & W                                                    & H                                                    & W                                               & H                        & W                        & H                        & W                        & H                        & W                        & H                        & \multicolumn{1}{c}{W}     & \multicolumn{1}{c}{H}     & W                                               & H                                              \\ \hline
S23                        & 0.270                                                & \textbf{0.131}                                       & \textbf{0.209}                                  & 0.193                    & 0.234                    & 0.202                    & 0.278                    & 0.211                    & 0.244                    & 0.208                    & \multicolumn{1}{c}{0.262} & \multicolumn{1}{c}{0.179} & 0.1059                                          & 0.0003                                         \\ \hline
S31                        & \textbf{0.193}                                       & \textbf{0.163}                                       & 0.321                                           & 0.185                    & 0.278                    & 0.223                    & 0.246                    & 0.286                    & 0.243                    & 0.216                    & \multicolumn{1}{c}{0.251} & \multicolumn{1}{c}{0.213} & 0.0070                                          & 0.0209                                         \\ \hline
S47                        & 0.218                                                & \textbf{0.099}                                       & 0.198                                           & 0.145                    & 0.204                    & 0.137                    & \textbf{0.149}           & 0.191                    & 0.188                    & 0.143                    & \multicolumn{1}{c}{0.168} & \multicolumn{1}{c}{0.194} & 0.0226                                          & 0.0073                                         \\ \hline
S48                        & \textbf{0.149}                                       & \textbf{0.071}                                       & 0.198                                           & 0.083                    & 0.184                    & 0.084                    & 0.164                    & 0.145                    & 0.157                    & 0.101                    & \multicolumn{1}{c}{0.156} & \multicolumn{1}{c}{0.096} & 0.0508                                          & 0.0532                                         \\ \hline
\end{tabular}}
\caption{The RMS error averaged over 4 sentences for width (W) and height (H) of the mouth of the real speakers classified under the low class of index 7 (mouth height), their corresponding 3D heads and the non-corresponding 3D heads. Values in bold means the lowest RMS error. The last column shows the p value of the t-test results between each corresponding 3D head and the non-corresponding 3D heads for width and height.}
\end{center}
\label{tab:Low_In7}
\end{table*}

\begin{figure}[ ]
\begin{center}
    \begin{footnotesize}{
    \begin{tabular}{m{0.30\columnwidth}m{0.13\columnwidth}m{0.13\columnwidth}m{0.13\columnwidth}}
        \makecell[l]{Real speaker\\ (ID: S47)}
        & \includegraphics[width=0.17\columnwidth]{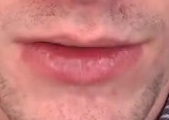}
        & \includegraphics[width=0.17\columnwidth]{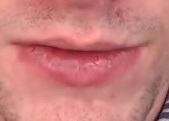}
        & \includegraphics[width=0.17\columnwidth]{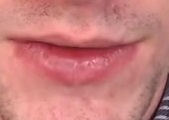}
        \\
        \makecell[l]{The corresponding \\ 3D head (ID: S47)}
        & \includegraphics[width=0.17\columnwidth]{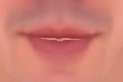}
        & \includegraphics[width=0.17\columnwidth]{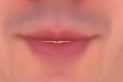}
        & \includegraphics[width=0.17\columnwidth]{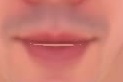}
        \\
        \makecell[l]{The non-corresponding \\ low 3D head (ID: S23)}
        & \includegraphics[width=0.17\columnwidth]{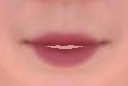}
        & \includegraphics[width=0.17\columnwidth]{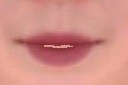}
        & \includegraphics[width=0.17\columnwidth]{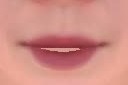}
        \\
        \makecell[l]{The non-corresponding \\ middle 3D head\\ (ID: S48)}
        & \includegraphics[width=0.17\columnwidth]{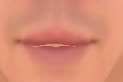}
        & \includegraphics[width=0.17\columnwidth]{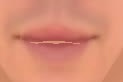}
        & \includegraphics[width=0.17\columnwidth]{figs/Mid_s47_s48_bwio7n_C_22}
        \\
        \makecell[l]{The non-corresponding \\ high 3D head \\(ID: S22)}
        & \includegraphics[width=0.17\columnwidth]{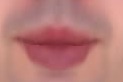}
        & \includegraphics[width=0.17\columnwidth]{figs/High_s47_s22_bwio7n_C_22.jpg}
        & \includegraphics[width=0.17\columnwidth]{figs/High_s47_s22_bwio7n_C_22.jpg}
        \end{tabular}
    }\end{footnotesize}
    \caption{Consecutive frames of the phoneme /ih/ during utterance of the word "in" from sentence "bin white in O seven now" for a real speaker (ID: S47) who is classified under the low class of index 7, the corresponding 3D head, the non-corresponding low, the non-corresponding middle and the non-corresponding high 3D heads.} 
    \label{fig:Low_Mid_High_In7}
\end{center}
\end{figure}

Figure \ref{fig:Low_Mid_High_In7} provides an example of consecutive frames of the phoneme /ih/ during the utterance of the word "in" from the sentence "bin white in O seven now" for a real speaker (ID: S47) who was classified in the low class of index 7, the corresponding 3D head, the non-corresponding low 3D head, the non-corresponding middle 3D head and the non-corresponding high 3D head. This figure illustrates how the non-corresponding high 3D head failed to detect the uttered phoneme and that the mouth was completely closed due to lip thickness, while the corresponding 3D head, the non-corresponding low 3D head and the non-corresponding middle 3D head gave the closest mouth shapes to the real speaker. These findings and the clear visual discrimination in the 3D lip motions presented by each 3D head suggest that an appropriate animation can be achieved by mapping between 2D videos of real speakers and the corresponding low 3D head, the non-corresponding low 3D heads or the non-corresponding middle 3D heads but not the non-corresponding high 3D heads. 

\begin{figure}
\begin{center}
    \includegraphics[trim = 13.6mm 43mm 08mm 10.9mm, clip,width=3.20in,keepaspectratio]{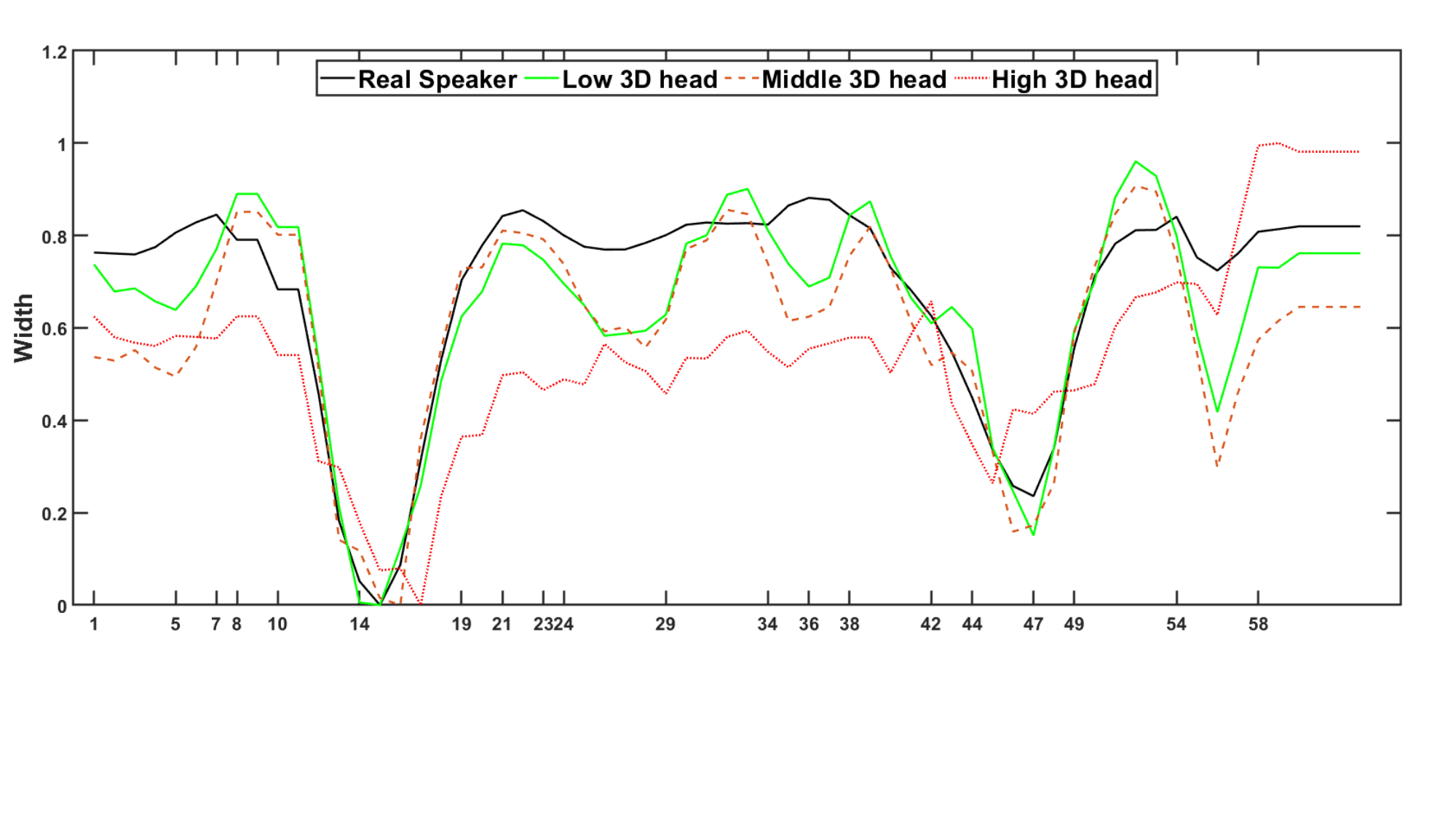}
    \includegraphics[trim =17.6mm 17mm 01mm 10mm, clip,width=3.20in,keepaspectratio]{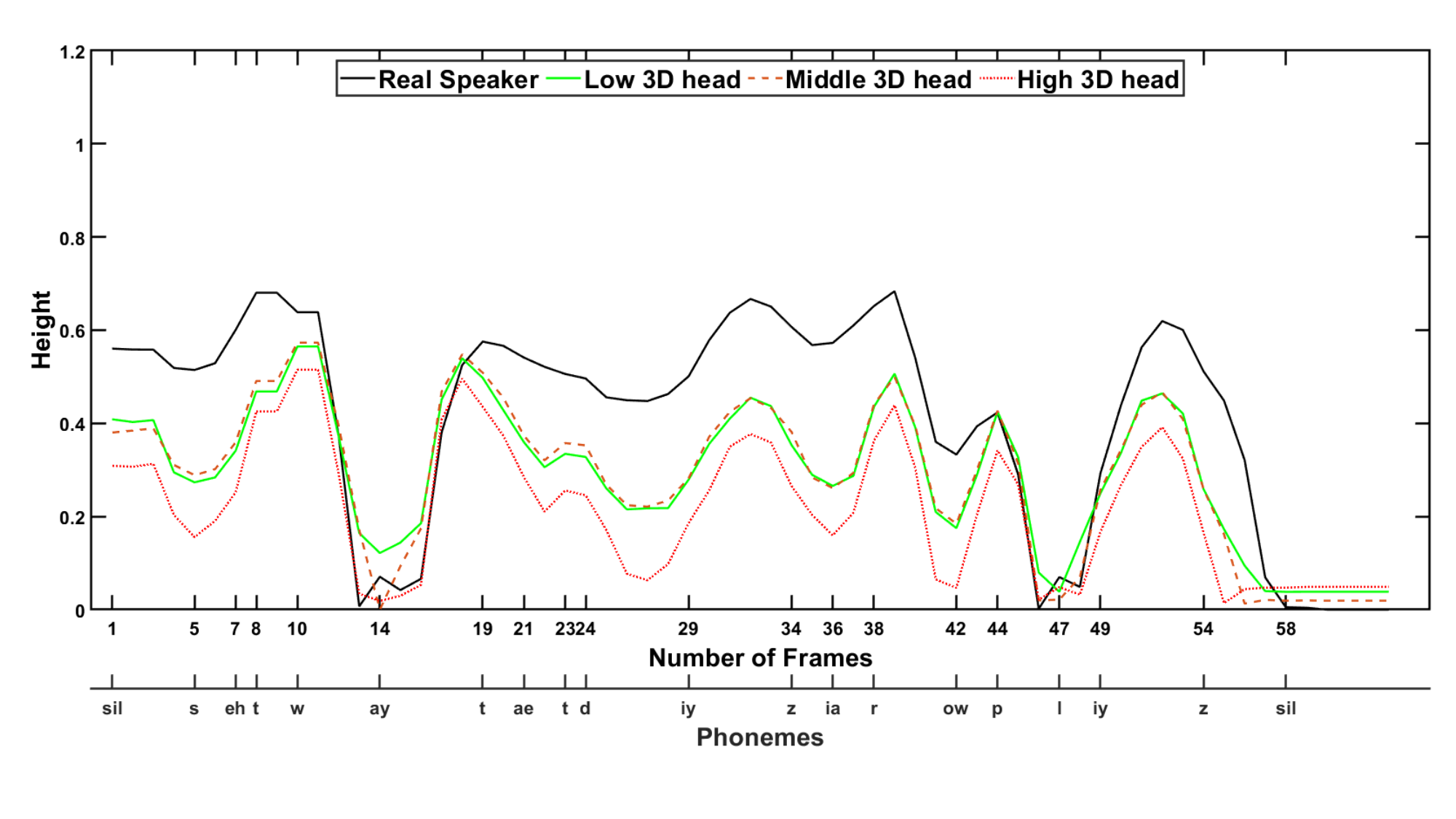}
    \caption{Width (upper) and height (lower) of mouth trajectories of 2D frames of the real speaker (ID: S31) classified under the low class of index 7, the corresponding 3D head, the non-corresponding middle 3D head (ID: S32) and the non-corresponding high 3D head (ID: S19), whilst uttering the sentence "set white at D zero please".}
    \label{fig:RMSEgraphs_Low_Mid_High}
\end{center}
\end{figure}

Figure \ref{fig:RMSEgraphs_Low_Mid_High} shows the trajectories of the width and the height parameters of the mouth aperture for the real speaker (ID: S31) classified under the low class of index 7, the corresponding low 3D head, the non-corresponding middle 3D head (ID: S32) and the non-corresponding high 3D head (ID: S19), whilst uttering the sentence "set white at D zero please". Whilst all the trajectories generated using the animation pipeline generally follow the real speaker’s trajectory, the trajectories of the corresponding 3D head and the non-corresponding middle 3D head are closer to the ground truth trajectories. For the width, the trajectory of the non-corresponding high 3D head shows a marked rise for the bilabial phoneme /p/, which confirms that the lips stretch due to touch between the upper and lower lips caused by the lip thickness. Also, for the height, the trajectory of this 3D head shows steep drops for dental phonemes such as /s/, /t/ and /d/, which confirms that the lips are semi-closed during uttering these phonemes. 

\begin{table*}[]
\begin{center}
\scalebox{0.80}{
\begin{tabular}{cccccccccccllcc}
\hline
 & \multicolumn{2}{c}{} & \multicolumn{10}{c}{\begin{tabular}[c]{@{}c@{}}Non-corresponding \\ 3D head (middle)\end{tabular}} & \multicolumn{2}{c}{} \\ \cline{4-13}
 & \multicolumn{2}{c}{\multirow{-2}{*}{\begin{tabular}[c]{@{}c@{}}Corresponding 3D \\ head (middle)\end{tabular}}} & \multicolumn{2}{c}{S32} & \multicolumn{2}{c}{S42} & \multicolumn{2}{c}{S54} & \multicolumn{2}{c}{S55} & \multicolumn{2}{l}{\cellcolor[HTML]{C0C0C0}} & \multicolumn{2}{c}{\multirow{-2}{*}{\begin{tabular}[c]{@{}c@{}}T-test\\ (P value)\end{tabular}}} \\ \cline{2-15} 
\multirow{-3}{*}{2D video} & W & H & W & H & W & H & W & H & W & H & \cellcolor[HTML]{C0C0C0} & \cellcolor[HTML]{C0C0C0} & W & H \\ \hline
S32 & 0.111 & \textbf{0.056} & \cellcolor[HTML]{C0C0C0}{\color[HTML]{C0C0C0} } & \cellcolor[HTML]{C0C0C0} & 0.122 & 0.089 & \textbf{0.106} & 0.112 & 0.174 & 0.059 & \cellcolor[HTML]{C0C0C0} & \cellcolor[HTML]{C0C0C0} & 0.3790 & 0.1837 \\ \hline
S42 & \textbf{0.198} & 0.083 & 0.261 & \textbf{0.078} & \cellcolor[HTML]{C0C0C0} & \cellcolor[HTML]{C0C0C0} & 0.248 & 0.138 & 0.220 & 0.088 & \cellcolor[HTML]{C0C0C0} & \cellcolor[HTML]{C0C0C0} & 0.0653 & 0.4274 \\ \hline
S54 & 0.281 & 0.132 & \textbf{0.255} & 0.152 & 0.263 & \textbf{0.130} & \cellcolor[HTML]{C0C0C0} & \cellcolor[HTML]{C0C0C0} & 0.256 & 0.149 & \cellcolor[HTML]{C0C0C0} & \cellcolor[HTML]{C0C0C0} & 0.0118 & 0.2324 \\ \hline
S55 & \textbf{0.113} & 0.107 & 0.154 & \textbf{0.105} & 0.126 & 0.118 & 0.134 & 0.117 & \cellcolor[HTML]{C0C0C0} & \cellcolor[HTML]{C0C0C0} & \cellcolor[HTML]{C0C0C0} & \cellcolor[HTML]{C0C0C0} & 0.0953 & 0.2687 \\ \hline
\multicolumn{1}{l}{} & \multicolumn{1}{l}{} & \multicolumn{1}{l}{} & \multicolumn{10}{c}{\begin{tabular}[c]{@{}c@{}}Non-corresponding \\ 3D head (low)\end{tabular}} & \multicolumn{1}{l}{} & \multicolumn{1}{l}{} \\ \cline{4-13}
\multicolumn{1}{l}{\multirow{-2}{*}{}} & \multicolumn{1}{l}{\multirow{-2}{*}{}} & \multicolumn{1}{l}{\multirow{-2}{*}{}} & \multicolumn{2}{c}{S23} & \multicolumn{2}{c}{S31} & \multicolumn{2}{c}{S47} & \multicolumn{2}{c}{S48} & \multicolumn{2}{l}{\cellcolor[HTML]{C0C0C0}} & \multicolumn{1}{l}{\multirow{-2}{*}{}} & \multicolumn{1}{l}{\multirow{-2}{*}{}} \\ \hline
\multicolumn{1}{l}{} & W & H & W & H & W & H & W & H & W & H & \cellcolor[HTML]{C0C0C0} & \cellcolor[HTML]{C0C0C0} & W & H \\ \hline
S32 & \textbf{0.111} & 0.056 & 0.119 & 0.068 & 0.123 & \textbf{0.047} & 0.144 & 0.075 & 0.163 & 0.089 & \cellcolor[HTML]{C0C0C0} & \cellcolor[HTML]{C0C0C0} & 0.0820 & 0.2141 \\ \hline
S42 & 0.198 & 0.083 & \textbf{0.185} & \textbf{0.075} & 0.224 & 0.125 & 0.255 & 0.092 & 0.208 & 0.130 & \cellcolor[HTML]{C0C0C0} & \cellcolor[HTML]{C0C0C0} & 0.2669 & 0.1870 \\ \hline
S54 & 0.281 & \textbf{0.132} & \textbf{0.249} & 0.151 & 0.265 & 0.180 & \textbf{0.249} & 0.158 & 0.273 & 0.133 & \cellcolor[HTML]{C0C0C0} & \cellcolor[HTML]{C0C0C0} & 0.0351 & 0.0943 \\ \hline
S55 & 0.113 & 0.107 & 0.119 & \textbf{0.096} & 0.119 & 0.111 & \textbf{0.112} & 0.108 & 0.122 & 0.104 & \cellcolor[HTML]{C0C0C0} & \cellcolor[HTML]{C0C0C0} & 0.0997 & 0.5385 \\ \hline
\multicolumn{1}{l}{} & \multicolumn{1}{l}{} & \multicolumn{1}{l}{} & \multicolumn{10}{c}{\begin{tabular}[c]{@{}c@{}}Non-corresponding\\ 3D head (high)\end{tabular}} & \multicolumn{1}{l}{} & \multicolumn{1}{l}{} \\ \cline{4-13}
\multicolumn{1}{l}{\multirow{-2}{*}{}} & \multicolumn{1}{l}{\multirow{-2}{*}{}} & \multicolumn{1}{l}{\multirow{-2}{*}{}} & \multicolumn{2}{c}{S17} & \multicolumn{2}{c}{S19} & \multicolumn{2}{c}{S22} & \multicolumn{2}{c}{S35} & \multicolumn{2}{c}{S46} & \multicolumn{1}{l}{\multirow{-2}{*}{}} & \multicolumn{1}{l}{\multirow{-2}{*}{}} \\ \hline
 & W & H & W & H & W & H & W & H & W & H & \multicolumn{1}{c}{W} & \multicolumn{1}{c}{H} & W & H \\ \hline
S32 & \textbf{0.111} & \textbf{0.056} & 0.185 & 0.132 & 0.145 & 0.102 & 0.135 & 0.193 & 0.176 & 0.145 & \multicolumn{1}{c}{0.161} & \multicolumn{1}{c}{0.123} & 0.0060 & 0.0055 \\ \hline
S42 & 0.198 & \textbf{0.083} & 0.205 & 0.140 & 0.209 & 0.090 & 0.200 & 0.171 & 0.229 & 0.156 & \multicolumn{1}{c}{\textbf{0.172}} & \multicolumn{1}{c}{0.098} & 0.6150 & 0.0394 \\ \hline
S54 & 0.281 & 0.132 & 0.226 & 0.148 & \textbf{0.209} & 0.147 & 0.241 & \textbf{0.129} & 0.267 & 0.136 & \multicolumn{1}{c}{0.131} & \multicolumn{1}{c}{0.194} & 0.0130 & 0.1931 \\ \hline
S55 & 0.113 & 0.107 & 0.102 & \textbf{0.103} & \textbf{0.097} & 0.111 & 0.116 & 0.114 & 0.099 & 0.127 & \multicolumn{1}{c}{0.116} & \multicolumn{1}{c}{0.110} & 0.0832 & 0.2022 \\ \hline
\end{tabular}}
\caption{The RMS error averaged over 4 sentences for width (W) and height (H) of the mouth of the real speakers classified under the middle class of index 7 (mouth height), their corresponding 3D heads and the non-corresponding 3D heads. Values in bold means the lowest RMS error. The last column shows p value of the t-test results between each corresponding 3D head and the non-corresponding 3D heads for width and height.}
\label{tab:middle_In7}
\end{center}
\end{table*}

Table \ref{tab:middle_In7} shows the RMSE results averaged over four sentences for the width and height of the mouth aperture of real speakers classified in the middle class of index 7 (mouth height), their corresponding middle 3D heads, non-corresponding low, non-corresponding middle and non-corresponding high 3D heads. This shows variations in the RMSE results for width due to variations in the mouth width of the real speakers.  T-test results suggested no significant difference in RMSE results for height for all speakers and for three out of four speakers for width. The corresponding 3D head of a real speaker (ID: S54) suggested a significant difference for the width; this may be due to a large mouth width (index 10). For the non-corresponding high 3D heads, the t-test results show a significant difference in the RMSE scores for width for the corresponding 3D head of the real speaker (ID: S32); this is probably due to the small mouth width.

\begin{figure}[ ]
\begin{center}
    \begin{footnotesize}{
    \begin{tabular}{m{0.30\columnwidth}m{0.13\columnwidth}m{0.13\columnwidth}m{0.13\columnwidth}}
        \makecell[l]{Real speaker \\(ID: S32)}
        & \includegraphics[width=0.17\columnwidth]{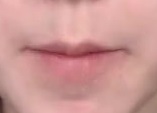}
        & \includegraphics[width=0.17\columnwidth]{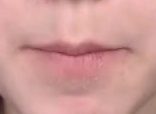}
        & \includegraphics[width=0.17\columnwidth]{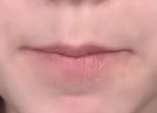}
        \\
        \makecell[l]{The corresponding \\ 3D head (ID: S32)}
        & \includegraphics[width=0.17\columnwidth]{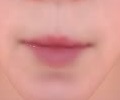}
        & \includegraphics[width=0.17\columnwidth]{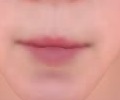}
        & \includegraphics[width=0.17\columnwidth]{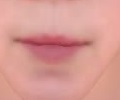}
       \\
        \makecell[l]{The non-corresponding \\ low 3D head \\(ID: S23)}
        & \includegraphics[width=0.17\columnwidth]{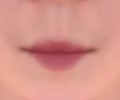}
        & \includegraphics[width=0.17\columnwidth]{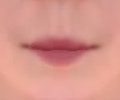}
        & \includegraphics[width=0.17\columnwidth]{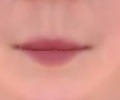}
        \\
         \makecell[l]{The non-corresponding \\ middle 3D head \\ (ID: S54)}
        & \includegraphics[width=0.17\columnwidth]{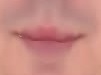}
        & \includegraphics[width=0.17\columnwidth]{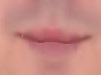}
        & \includegraphics[width=0.17\columnwidth]{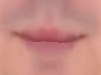}
        \\
        \makecell[l]{The non-corresponding \\ high 3D head\\ (ID: S19)}
        & \includegraphics[width=0.17\columnwidth]{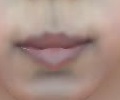}
        & \includegraphics[width=0.17\columnwidth]{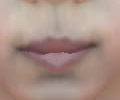}
        & \includegraphics[width=0.17\columnwidth]{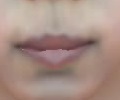}
        \end{tabular}
    }\end{footnotesize}
    \caption{Consecutive frames of the phoneme /b/ during utterance of the word "bin" from sentence "bin white at U three again" for a real speaker (ID: S32) who classified under the middle class of index 7, the corresponding 3D head, the non-corresponding low 3D head, the non-corresponding middle 3D head and the non-corresponding high 3D head.} 
    \label{fig:Mid_Low_High_In7}
\end{center}
\end{figure}

For the height, t-test results showed a significant difference for two of the corresponding middle 3D heads; this may be because their corresponding real speakers (IDs: S32 and S42) were classified in the low class and the low to middle class of index 10, respectively. This makes the mouth of the non-corresponding high 3D heads shrink to fit the real speakers’ mouths; thus, the lips are not closed or opened adequately.  Figure \ref{fig:Mid_Low_High_In7} confirms these findings by showing an example of consecutive frames  of the phoneme /b/ during the utterance of the word "bin" from the phrase "bin white at U three again" for a real speaker (ID: S32) classified in the middle class of index 7, the corresponding 3D head, the non-corresponding low 3D heads and the non-corresponding high 3D heads. These findings may confirm that the resulting 3D lip motions become sufficient and adequate when 2D videos of the real speakers classified in the middle class of index 7 are mapped to the corresponding 3D head, the non-corresponding middle 3D heads and the non-corresponding low 3D heads and when they are mapped to the non-corresponding high 3D heads that relate to real speakers who have a similar mouth width.

Table \ref{tab:high_In7} shows the RMSE results averaged over four sentences for the width and height of the mouth aperture of real speakers classified in the high class of index 7 (mouth height), their corresponding 3D heads, non-corresponding low, non-corresponding middle and non-corresponding high 3D heads. It can be observed that the corresponding 3D head of a real speaker (ID: S19) gave the lowest RMSE scores for width for real speakers (IDs: S22 and S46); this may due to similarities in mouth width. Additionally, the non-corresponding low 3D heads (IDs: S23, S47 and S
48) gave the lowest scores because their corresponding real speakers were classified in the middle class and the middle to high class of index 10, and most of the real speakers (IDs: S17, S19, S22 and S46) were classified in the middle to high class or the high class. This was confirmed by t-test results that showed no significant difference for width between the corresponding 3D heads and the non-corresponding low 3D heads.

For the height, the corresponding high 3D heads gave the lowest score for most of the speakers from different classes. T-test results showed no significant difference in the RMSE scores for height between four out of five of the corresponding high 3D heads and the non-corresponding middle 3D heads. However, there was a significant difference between three out of five of the corresponding high 3D heads and the non-corresponding low 3D heads. 

\begin{figure}[ ]
\begin{center}
    \begin{footnotesize}{
    \begin{tabular}{m{0.30\columnwidth}m{0.13\columnwidth}m{0.13\columnwidth}m{0.13\columnwidth}}
        \makecell[l]{Real speaker \\ (ID: S22)}
        & \includegraphics[width=0.17\columnwidth ]{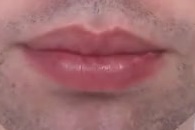}
        & \includegraphics[width=0.17\columnwidth]{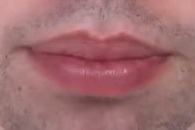}
        & \includegraphics[width=0.17\columnwidth]{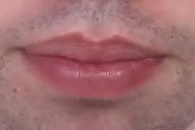}
        \\
        \makecell[l]{The corresponding \\ 3D head (ID: S22)}
        & \includegraphics[width=0.17\columnwidth]{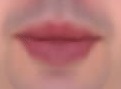}
        & \includegraphics[width=0.17\columnwidth]{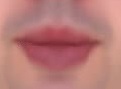}
        & \includegraphics[width=0.17\columnwidth]{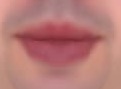}
        \\
        \makecell[l]{The non-corresponding \\ low 3D head \\ (ID: S47)}
        & \includegraphics[width=0.17\columnwidth]{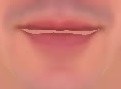}
        & \includegraphics[width=0.17\columnwidth]{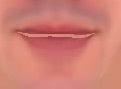}
        & \includegraphics[width=0.17\columnwidth]{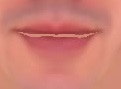}
        \\
        \makecell[l]{The non-corresponding \\ middle 3D head \\ (ID: S55)}
        & \includegraphics[width=0.17\columnwidth]{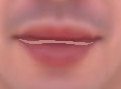}
        & \includegraphics[width=0.17\columnwidth]{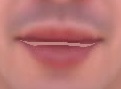}
        & \includegraphics[width=0.17\columnwidth]{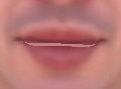}
         \\
        \makecell[l]{The non-corresponding \\ middle 3D head \\ (ID: S54)}
        & \includegraphics[width=0.17\columnwidth]{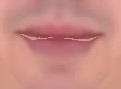}
        & \includegraphics[width=0.17\columnwidth]{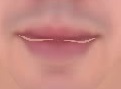}
        & \includegraphics[width=0.17\columnwidth]{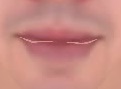}
        \end{tabular}
    }\end{footnotesize}
    \caption{Consecutive frames of the phoneme /b/ during utterance of the word "bin" from sentence "bin green by Q zero again" for a real speaker (ID: S22) who classified under the high class of index 7 (first row), the corresponding 3D head (second row), the non-corresponding low 3D head (third row) and the non-corresponding middle 3D head (last row).} 
    \label{fig:High_Low_Mid_In7_S22}
\end{center}

\end{figure}

Based on these findings, it can be concluded that mapping between 2D videos of real speakers classified in the high class of index 7 and the non-corresponding low 3D heads cannot achieve any reasonable 3D lip animations. While it is possible to achieve reasonable 3D lip motion using 2D videos of real speakers who have a middle mouth height, to animate 3D heads corresponding to real speakers who have a high mouth height, they must be similar in other facial features, such as lower or upper lip thickness (indices 8 and 9, respectively), mouth width (index 10) or the distance between the nose and the upper lip (index 12). This is indicated by Figure \ref{fig:High_Low_Mid_In7_S22} which shows how the lips of the non-corresponding 3D heads fail to give the mouth shape of the phoneme /b/. This figure also shows how the non-corresponding middle 3D head (ID: S54) gives a semi-opened mouth shape due to its high mouth width, which is similar to that of the real speaker (ID: S22). However, it fails to deliver the correct mouth shape due to its middle upper and lower lip thickness.

\begin{figure}[ ]
\begin{center}
    \begin{footnotesize}{
    \begin{tabular}{m{0.30\columnwidth}m{0.13\columnwidth}m{0.13\columnwidth}}
        \makecell[l]{Real speaker \\ (ID: S46)}
        & \includegraphics[width=0.17\columnwidth]{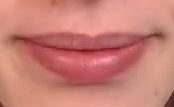}
        & \includegraphics[width=0.17\columnwidth]{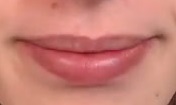}
        \\
        \makecell[l]{The corresponding \\ 3D head (ID: S46)}
        & \includegraphics[width=0.17\columnwidth]{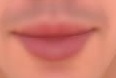}
        & \includegraphics[width=0.17\columnwidth]{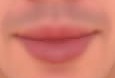}
        \\
        \makecell[l]{The non-corresponding \\ middle 3D head \\ (ID: S54)}
        & \includegraphics[width=0.17\columnwidth]{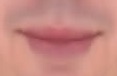}
        & \includegraphics[width=0.17\columnwidth]{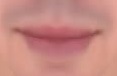}
         \\
        \makecell[l]{The non-corresponding \\ low 3D head \\ (ID: S48)}
        & \includegraphics[width=0.17\columnwidth]{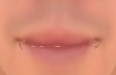}
        & \includegraphics[width=0.17\columnwidth]{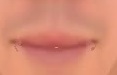}
        \\
        \makecell[l]{The non-corresponding \\ low 3D head \\ (ID: S31)}
        & \includegraphics[width=0.17\columnwidth]{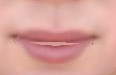}
        & \includegraphics[width=0.17\columnwidth]{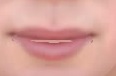}
        \end{tabular}
    }\end{footnotesize}
    \caption{Consecutive frames of the phoneme /p/ during utterance of the word "please" from sentence "lay white with A 5 please" for a real speaker (ID: S46) who classified under the high class of index 7 (first row), the corresponding 3D head (second row), the non-corresponding middle 3D head (third row) and the non-corresponding low 3D heads (last two rows).} 
    \label{fig:High_Low_Mid_In7_S46}
\end{center}
\end{figure}

Figure \ref{fig:High_Low_Mid_In7_S46} shows an example of consecutive frames of the phoneme /p/ during utterance of the word "please" from the phrase "lay white with A 5 please" for a real speaker (ID: S46) classified in the high class of index 7, the corresponding 3D head, the non-corresponding middle 3D heads and the non-corresponding low 3D heads. This figure illustrates how the non-corresponding middle 3D (ID: S54) heads gave a mouth shape more similar to that of the real speaker due to similarities in mouth width (index 10) and middle lower lip thickness. The non-corresponding low 3D heads (ID: S48) gave a more accurate mouth shape (semi-closed mouth shape) because of the large distance between the nose and the upper lip (index 12), the middle mouth width (index 10) and the middle upper and lower lip thickness (indices 8 and 9), while the non-corresponding low 3D heads (ID: S31) failed to give the correct mouth shape due to the low to middle mouth width and upper and lower lip thickness. The distortion in the texture around the mouth’s corners of the non-corresponding low 3D heads (IDs: S48 and S31) was due to differences in mouth width between the real speaker and the 3D heads.

\begin{figure}
\begin{center}
    \includegraphics[trim = 14mm 45mm 08mm 10.5mm, clip,width=3.25in,keepaspectratio]{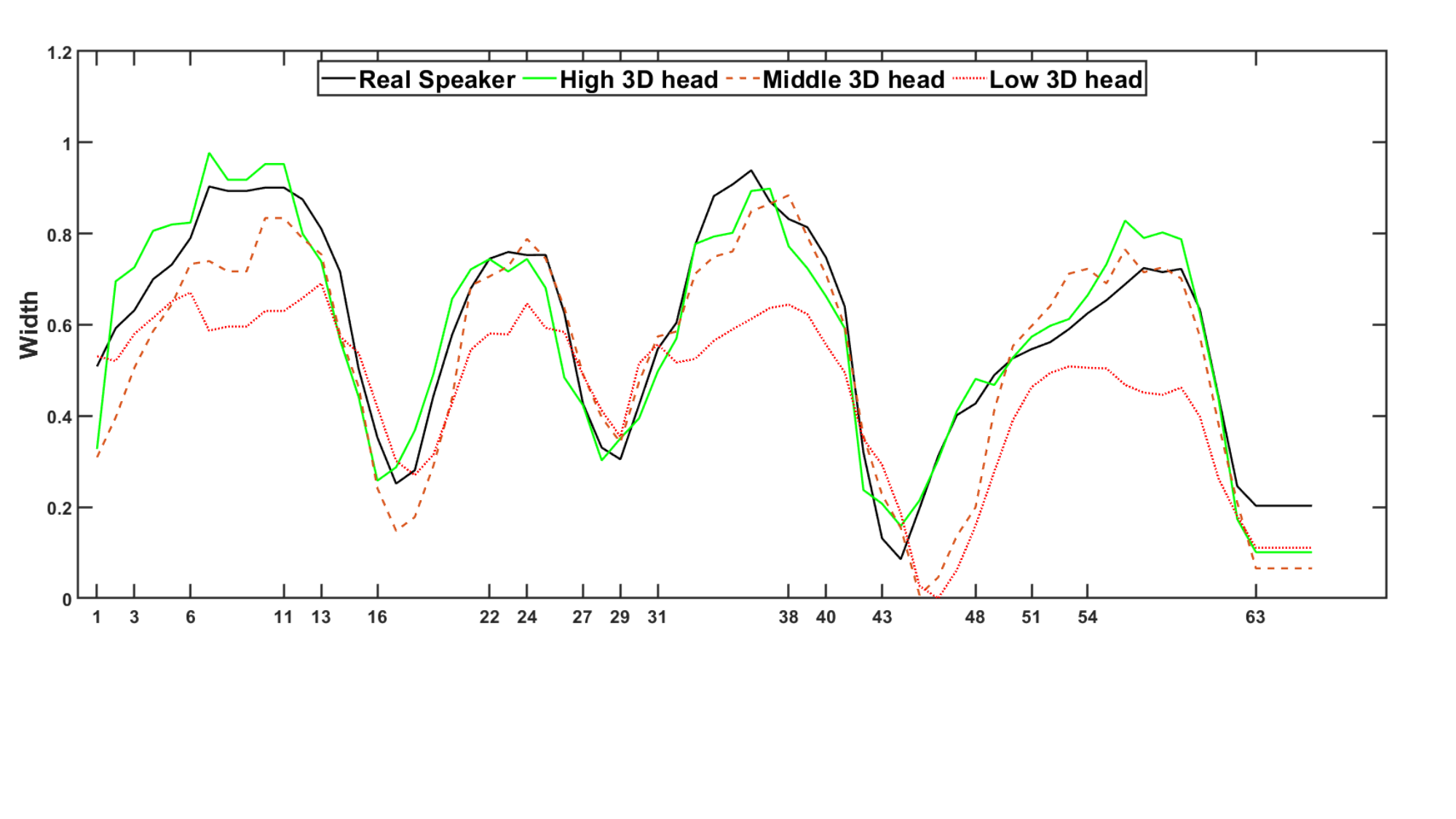}
    \includegraphics[trim = 18mm 22mm 10mm 10mm, clip,width=3.25in,keepaspectratio]{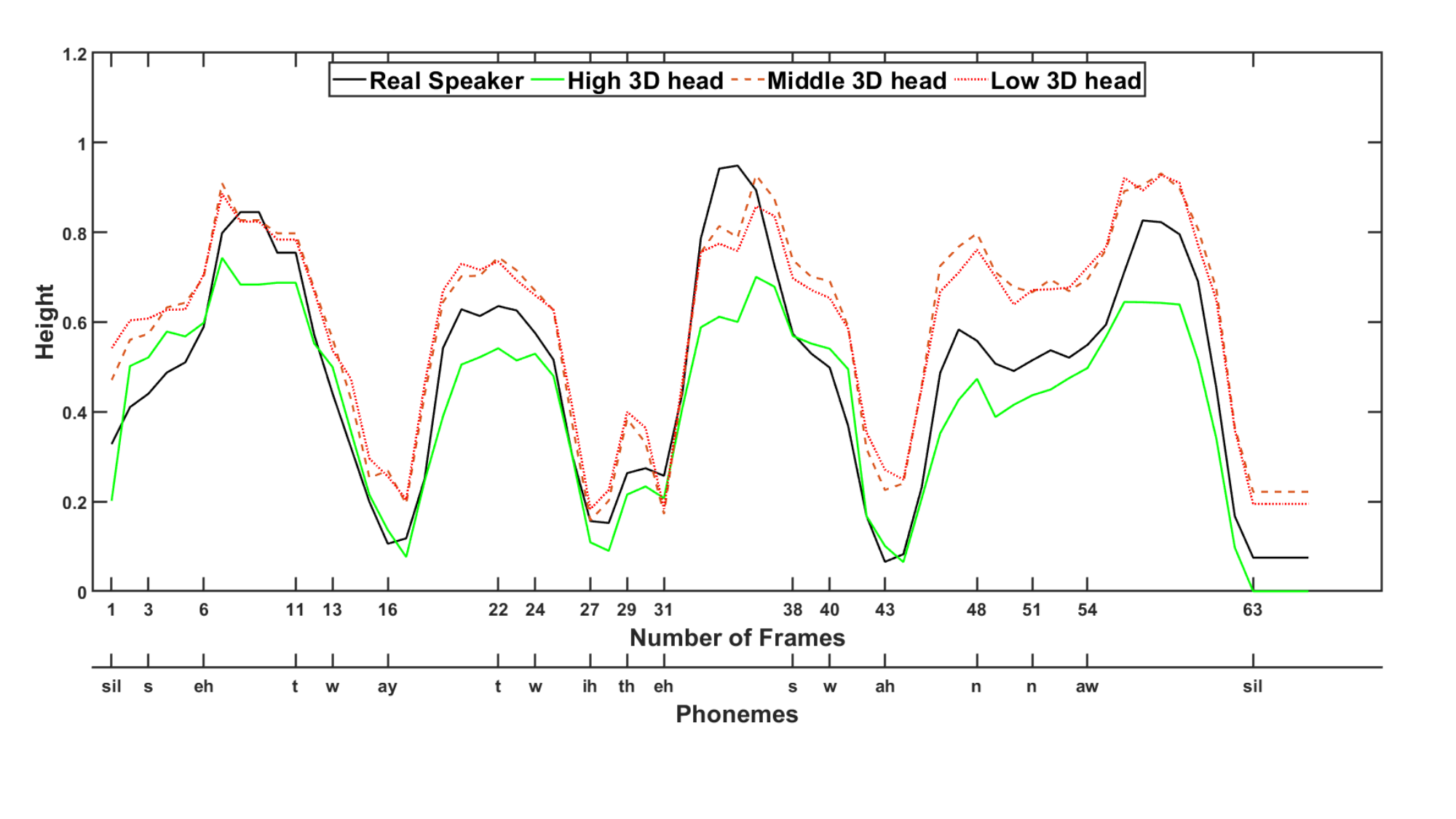}
    \caption{Width (upper) and height (lower) of mouth trajectories of 2D frames of the real speaker (ID: S22) classified under the high class of index 7, the corresponding 3D head, the non-corresponding middle 3D head (ID: S55) and the non-corresponding low 3D head (ID: S48), whilst uttering the sentence "set white with S 1 now".}
    \label{fig:RMSEgraphs_High_Mid_Low}
\end{center}
\end{figure}

Figure \ref{fig:RMSEgraphs_High_Mid_Low} shows the trajectories of the width and the height parameters of the mouth aperture for the real speaker (ID: S22) classified under the high class of index 7, the corresponding 3D head, the non-corresponding middle 3D head (ID: S55) and the non-corresponding low 3D head (ID: S48), whilst uttering the sentence "set white with S 1 now". The trajectories of the corresponding 3D head are closer to the ground truth trajectories for both width and height. For the width, what can be clearly seen in this figure is the steady decline of the trajectories of the non-corresponding low 3D head. For the height, the trajectories of the non-corresponding low 3D head show a marked increase for the rounding lips phonemes such as /w/ and /aw/, alveolar phonemes such as /th/, and dental phonemes such as /s/, /t/, and /n/, which confirms that the lips are widely opened during uttering these phonemes that require semi-opened mouth shape. 


\begin{table*}[]
\begin{center}
\scalebox{0.80}{
\begin{tabular}{cccccccccccllcc}
\hline
 & \multicolumn{2}{c}{} & \multicolumn{10}{c}{\begin{tabular}[c]{@{}c@{}}Non-corresponding \\ 3D head (high)\end{tabular}} & \multicolumn{2}{c}{} \\ \cline{4-13}
 & \multicolumn{2}{c}{\multirow{-2}{*}{\begin{tabular}[c]{@{}c@{}}Corresponding 3D \\ head (high)\end{tabular}}} & \multicolumn{2}{c}{S17} & \multicolumn{2}{c}{S19} & \multicolumn{2}{c}{S22} & \multicolumn{2}{c}{S35} & \multicolumn{2}{c}{S46} & \multicolumn{2}{c}{\multirow{-2}{*}{\begin{tabular}[c]{@{}c@{}}T-test\\ (P value)\end{tabular}}} \\ \cline{2-15} 
\multirow{-3}{*}{2D video} & W & H & W & H & W & H & W & H & W & H &  &  & W & H \\ \hline
S17 & \textbf{0.092} & \textbf{0.095} & \cellcolor[HTML]{C0C0C0}{\color[HTML]{C0C0C0} } & \cellcolor[HTML]{C0C0C0} & 0.101 & 0.130 & 0.112 & 0.113 & 0.107 & 0.111 & 0.111 & 0.110 & 0.0080 & 0.0210 \\ \hline
S19 & \textbf{0.215} & 0.122 & 0.259 & 0.132 & \cellcolor[HTML]{C0C0C0} & \cellcolor[HTML]{C0C0C0} & 0.222 & 0.134 & 0.250 & \textbf{0.120} & 0.247 & 0.128 & 0.0337 & 0.1266 \\ \hline
S22 & 0.172 & \textbf{0.137} & 0.136 & 0.150 & \textbf{0.125} & 0.184 & \cellcolor[HTML]{C0C0C0} & \cellcolor[HTML]{C0C0C0} & 0.127 & 0.157 & 0.149 & 0.156 & 0.0062 & 0.0469 \\ \hline
S35 & 0.311 & \textbf{0.149} & 0.270 & 0.181 & 0.294 & 0.152 & \textbf{0.261} & 0.186 & \cellcolor[HTML]{C0C0C0} & \cellcolor[HTML]{C0C0C0} & 0.499 & 0.226 & 0.7465 & 0.0919 \\ \hline
S46 & 0.231 & 0.118 & \multicolumn{1}{l}{0.198} & \multicolumn{1}{l}{0.142} & \multicolumn{1}{l}{\textbf{0.166}} & \multicolumn{1}{l}{0.136} & \multicolumn{1}{l}{0.227} & \multicolumn{1}{l}{\textbf{0.070}} & \multicolumn{1}{l}{0.244} & \multicolumn{1}{l}{0.108} & \cellcolor[HTML]{C0C0C0} & \cellcolor[HTML]{C0C0C0} & \multicolumn{1}{l}{0.2847} & \multicolumn{1}{l}{0.8234} \\ \hline
\multicolumn{1}{l}{} & \multicolumn{1}{l}{} & \multicolumn{1}{l}{} & \multicolumn{10}{c}{\begin{tabular}[c]{@{}c@{}}Non-corresponding \\ 3D head (low)\end{tabular}} & \multicolumn{1}{l}{} & \multicolumn{1}{l}{} \\ \cline{4-13}
\multicolumn{1}{l}{\multirow{-2}{*}{}} & \multicolumn{1}{l}{\multirow{-2}{*}{}} & \multicolumn{1}{l}{\multirow{-2}{*}{}} & \multicolumn{2}{c}{S23} & \multicolumn{2}{c}{S31} & \multicolumn{2}{c}{S47} & \multicolumn{2}{c}{S48} & \multicolumn{2}{l}{\cellcolor[HTML]{C0C0C0}} & \multicolumn{1}{l}{\multirow{-2}{*}{}} & \multicolumn{1}{l}{\multirow{-2}{*}{}} \\ \hline
\multicolumn{1}{l}{} & W & H & W & H & W & H & W & H & W & H & \cellcolor[HTML]{C0C0C0} & \cellcolor[HTML]{C0C0C0} & W & H \\ \hline
S17 & 0.092 & \textbf{0.095} & \textbf{0.087} & 0.110 & 0.097 & 0.116 & 0.112 & 0.128 & 0.111 & 0.110 & \cellcolor[HTML]{C0C0C0} & \cellcolor[HTML]{C0C0C0} & 0.2021 & 0.0158 \\ \hline
S19 & 0.215 & 0.122 & 0.240 & 0.122 & 0.224 & 0.125 & 0.254 & \textbf{0.079} & \textbf{0.214} & 0.130 & \cellcolor[HTML]{C0C0C0} & \cellcolor[HTML]{C0C0C0} & 0.1337 & 0.5459 \\ \hline
S22 & 0.172 & \textbf{0.137} & 0.126 & 0.149 & \textbf{0.111} & 0.160 & 0.139 & 0.142 & 0.179 & 0.153 & \cellcolor[HTML]{C0C0C0} & \cellcolor[HTML]{C0C0C0} & 0.1070 & 0.0338 \\ \hline
S35 & 0.311 & 0.149 & 0.315 & \textbf{0.128} & 0.320 & 0.115 & 0.319 & 0.149 & \textbf{0.302} & 0.156 & \cellcolor[HTML]{C0C0C0} & \cellcolor[HTML]{C0C0C0} & 0.5214 & 0.2934 \\ \hline
S46 & 0.231 & \textbf{0.118} & \multicolumn{1}{l}{0.267} & \multicolumn{1}{l}{0.148} & \multicolumn{1}{l}{0.237} & \multicolumn{1}{l}{0.166} & \multicolumn{1}{l}{0.267} & \multicolumn{1}{l}{0.137} & \multicolumn{1}{l}{\textbf{0.215}} & \multicolumn{1}{l}{0.138} & \cellcolor[HTML]{C0C0C0} & \cellcolor[HTML]{C0C0C0} & \multicolumn{1}{l}{0.3082} & \multicolumn{1}{l}{0.0224} \\ \hline
\multicolumn{1}{l}{} & \multicolumn{1}{l}{} & \multicolumn{1}{l}{} & \multicolumn{10}{c}{\begin{tabular}[c]{@{}c@{}}Non-corresponding\\ 3D head (middle)\end{tabular}} & \multicolumn{1}{l}{} & \multicolumn{1}{l}{} \\ \cline{4-13}
\multicolumn{1}{l}{\multirow{-2}{*}{}} & \multicolumn{1}{l}{\multirow{-2}{*}{}} & \multicolumn{1}{l}{\multirow{-2}{*}{}} & \multicolumn{2}{c}{S32} & \multicolumn{2}{c}{S42} & \multicolumn{2}{c}{S54} & \multicolumn{2}{c}{S55} & \multicolumn{2}{c}{\cellcolor[HTML]{C0C0C0}} & \multicolumn{1}{l}{\multirow{-2}{*}{}} & \multicolumn{1}{l}{\multirow{-2}{*}{}} \\ \hline
 & W & H & W & H & W & H & W & H & W & H & \multicolumn{1}{c}{\cellcolor[HTML]{C0C0C0}} & \multicolumn{1}{c}{\cellcolor[HTML]{C0C0C0}} & W & H \\ \hline
S17 & \textbf{0.092} & \textbf{0.095} & 0.120 & 0.112 & 0.112 & 0.126 & 0.107 & 0.122 & 0.109 & 0.113 & \multicolumn{1}{c}{\cellcolor[HTML]{C0C0C0}} & \multicolumn{1}{c}{\cellcolor[HTML]{C0C0C0}} & 0.0060 & 0.0065 \\ \hline
S19 & \textbf{0.215} & \textbf{0.122} & 0.262 & 0.125 & 0.224 & 0.127 & 0.238 & 0.140 & 0.281 & 0.131 & \multicolumn{1}{c}{\cellcolor[HTML]{C0C0C0}\textbf{}} & \multicolumn{1}{c}{\cellcolor[HTML]{C0C0C0}} & 0.0642 & 0.0783 \\ \hline
S22 & 0.172 & \textbf{0.137} & 0.300 & 0.233 & \textbf{0.132} & 0.149 & 0.147 & 0.156 & 0.168 & 0.157 & \multicolumn{1}{c}{\cellcolor[HTML]{C0C0C0}} & \multicolumn{1}{c}{\cellcolor[HTML]{C0C0C0}} & 0.7269 & 0.1609 \\ \hline
S35 & 0.311 & 0.149 & 0.325 & \textbf{0.130} & 0.325 & 0.153 & \textbf{0.293} & 0.156 & 0.313 & 0.141 & \multicolumn{1}{c}{\cellcolor[HTML]{C0C0C0}} & \multicolumn{1}{c}{\cellcolor[HTML]{C0C0C0}} & 0.7177 & 0.5501 \\ \hline
S46 & \textbf{0.231} & 0.118 & \multicolumn{1}{l}{0.269} & \multicolumn{1}{l}{0.191} & \multicolumn{1}{l}{0.245} & \multicolumn{1}{l}{0.114} & \multicolumn{1}{l}{0.255} & \multicolumn{1}{l}{\textbf{0.104}} & \multicolumn{1}{l}{0.252} & \multicolumn{1}{l}{0.144} & \cellcolor[HTML]{C0C0C0} & \cellcolor[HTML]{C0C0C0} & \multicolumn{1}{l}{0.0171} & \multicolumn{1}{l}{0.3760}
\end{tabular}}
\caption{The RMS error averaged over 4 sentences for width (W) and height (H) of the mouth of the real speakers classified under the high class of index 7 (mouth height), their corresponding 3D heads and the non-corresponding 3D heads. Values in bold means the lowest RMS error. The last column shows p value of the t-test results between each corresponding 3D head and the non-corresponding 3D heads for width and height.}
\label{tab:high_In7}
\end{center}
\end{table*}


\subsection{Mouth width (index 10)}
\label{subsec:In10}

Table \ref{tab:low_In10} shows the RMSE results averaged over four sentences for width and height of the mouth aperture of real speakers classified in the low class of index 10 (mouth width), their corresponding 3D heads, non-corresponding low, non-corresponding middle and non-corresponding high 3D heads. From this table, it can be observed that the 3D head that corresponded to a real speaker (ID: S32) gave the lowest RMSE score for three out of four speakers for width and for all speakers for height, when it was fitted to 2D videos of real speakers who were classified under the low class. This may be due to its middle mouth height (i.e. index 7). This explains why it failed to give the lowest score for width for a real speaker (ID: S35) with a high mouth height.


\begin{table*}[htb]
\resizebox{\linewidth}{!}{
\begin{tabular}{cccccccccccllcc}
\hline
 & \multicolumn{2}{c}{} & \multicolumn{10}{c}{\begin{tabular}[c]{@{}c@{}}Non-corresponding \\ 3D head (low)\end{tabular}} & \multicolumn{2}{c}{} \\ \cline{4-13}
 & \multicolumn{2}{c}{{\begin{tabular}[c]{@{}c@{}}Corresponding 3D \\ head (low)\end{tabular}}} & \multicolumn{2}{c}{S16} & \multicolumn{2}{c}{S32} & \multicolumn{2}{c}{S35} & \multicolumn{2}{c}{S38} & \multicolumn{2}{l}{\cellcolor[HTML]{C0C0C0}} & \multicolumn{2}{c}{{\begin{tabular}[c]{@{}c@{}}T-test\\ (P value)\end{tabular}}} \\ \cline{2-15} 
{2D video} & W & H & W & H & W & H & W & H & W & H & \cellcolor[HTML]{C0C0C0} & \cellcolor[HTML]{C0C0C0} & W & H \\ \hline
S16 & 0.173 & \textbf{0.097} & \cellcolor[HTML]{C0C0C0}{\color[HTML]{C0C0C0} } & \cellcolor[HTML]{C0C0C0} & \textbf{0.162} & \textbf{0.097} & 0.176 & 0.108 & 0.191 & 0.149 & \cellcolor[HTML]{C0C0C0} & \cellcolor[HTML]{C0C0C0} & 0.7290 & 0.3157 \\ \hline
S32 & \textbf{0.111} & \textbf{0.056} & 0.135 & 0.118 & \cellcolor[HTML]{C0C0C0} & \cellcolor[HTML]{C0C0C0} & 0.176 & 0.145 & 0.140 & 0.138 & \cellcolor[HTML]{C0C0C0} & \cellcolor[HTML]{C0C0C0} & 0.0930 & 0.0107 \\ \hline
S35 & 0.311 & 0.149 & \textbf{0.258} & 0.143 & 0.325 & \textbf{0.130} & \cellcolor[HTML]{C0C0C0} & \cellcolor[HTML]{C0C0C0} & 0.306 & 0.167 & \cellcolor[HTML]{C0C0C0} & \cellcolor[HTML]{C0C0C0} & 0.5385 & 0.8495 \\ \hline
S38 & 0.296 & 0.138 & 0.273 & 0.147 & \textbf{0.204} & \textbf{0.111} & 0.214 & 0.121 & \cellcolor[HTML]{C0C0C0} & \cellcolor[HTML]{C0C0C0} & \cellcolor[HTML]{C0C0C0} & \cellcolor[HTML]{C0C0C0} & 0.0928 & 0.3905 \\ \hline
\multicolumn{1}{l}{} & \multicolumn{1}{l}{} & \multicolumn{1}{l}{} & \multicolumn{10}{c}{\begin{tabular}[c]{@{}c@{}}Non-corresponding \\ 3D head (middle)\end{tabular}} & \multicolumn{1}{l}{} & \multicolumn{1}{l}{} \\ \cline{4-13}
\multicolumn{1}{l}{} & \multicolumn{1}{l}{} & \multicolumn{1}{l}{} & \multicolumn{2}{c}{S7} & \multicolumn{2}{c}{S15} & \multicolumn{2}{c}{S24} & \multicolumn{2}{c}{S48} & \multicolumn{2}{c}{S55} & \multicolumn{1}{l}{} & \multicolumn{1}{l}{} \\ \hline
\multicolumn{1}{l}{} & W & H & W & H & W & H & W & H & W & H & W & H & W & H \\ \hline
S16 & 0.173 & 0.097 & \textbf{0.127} & 0.120 & 0.137 & 0.100 & 0.154 & \textbf{0.093} & 0.212 & 0.096 & 0.181 & 0.103 & 0.5231 & 0.3162 \\ \hline
S32 & \textbf{0.111} & \textbf{0.056} & 0.117 & 0.198 & 0.133 & 0.085 & 0.152 & 0.103 & 0.163 & 0.089 & 0.175 & 0.059 & 0.0235 & 0.1006 \\ \hline
S35 & 0.311 & 0.149 & 0.277 & 0.200 & \textbf{0.274} & \textbf{0.140} & 0.334 & 0.145 & 0.302 & 0.156 & 0.313 & 0.141 & 0.3838 & 0.5471 \\ \hline
S38 & 0.296 & 0.138 & 0.300 & 0.183 & 0.287 & \textbf{0.112} & 0.247 & 0.117 & \textbf{0.174}& 0.119 & 0.212 & 0.114 & 0.0901 & 0.5430 \\ \hline
\multicolumn{1}{l}{} & \multicolumn{1}{l}{} & \multicolumn{1}{l}{} & \multicolumn{10}{c}{\begin{tabular}[c]{@{}c@{}}Non-corresponding\\ 3D head (high)\end{tabular}} & \multicolumn{1}{l}{} & \multicolumn{1}{l}{} \\ \cline{4-13}
\multicolumn{1}{l}{} & \multicolumn{1}{l}{} & \multicolumn{1}{l}{} & \multicolumn{2}{c}{S19} & \multicolumn{2}{c}{S20} & \multicolumn{2}{c}{S22} & \multicolumn{2}{c}{S46} & \multicolumn{2}{c}{S54} & \multicolumn{1}{l}{} & \multicolumn{1}{l}{} \\ \hline
 & W & H & W & H & W & H & W & H & W & H & \multicolumn{1}{c}{W} & \multicolumn{1}{c}{H} & W & H \\ \hline
S16 & 0.173 & 0.097 & 0.179 & 0.095 & \textbf{0.126} & 0.127 & 0.145 & 0.103 & 0.181 & \textbf{0.094} & \multicolumn{1}{c}{0.213} & \multicolumn{1}{c}{0.116} & 0.7957 & 0.1912 \\ \hline
S32 & 0.111 & \textbf{0.056} & 0.145 & 0.102 & 0.134 & 0.072 & 0.135 & 0.193 & 0.161 & 0.123 & \multicolumn{1}{c}{\textbf{0.106}} & \multicolumn{1}{c}{0.112} & 0.0484 & 0.0325 \\ \hline
S35 & 0.311 & 0.149 & 0.294 & 0.152 & 0.288 & \textbf{0.132} & \textbf{0.261} & 0.186 & 0.499 & 0.226 & \multicolumn{1}{c}{0.293} & \multicolumn{1}{c}{0.156} & 0.7312 & 0.2610 \\ \hline
S38 & 0.296 & 0.138 & 0.269 & 0.111 & 0.254 & 0.111 & 0.254 & \textbf{0.105} & 0.229 & 0.145 & \multicolumn{1}{c}{\textbf{0.199}} & \multicolumn{1}{c}{0.123} & 0.0111 & 0.0561 \\ \hline
\end{tabular}}
\caption{The RMS error averaged over 4 sentences for width (W) and height (H) of the mouth of the real speakers classified under the low class of index 10 (mouth width), their corresponding 3D heads and the non-corresponding 3D heads. Values in bold means the lowest RMS error. The last column shows p value of the t-test results between each corresponding 3D head and the non-corresponding 3D heads for width and height.}
\label{tab:low_In10}
\end{table*}

\begin{figure}[htb]
\begin{center}
    \begin{footnotesize}{
    \resizebox{0.8\linewidth}{!}{
    \begin{tabular}{m{0.30\columnwidth}m{0.17\columnwidth}m{0.17\columnwidth}m{0.17\columnwidth}}
        \makecell[l]{Real speaker\\ (ID: S16)}
        & \includegraphics[width=0.19\columnwidth]{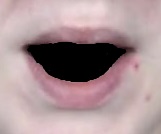}
        & \includegraphics[width=0.19\columnwidth]{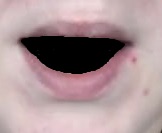}
        & \includegraphics[width=0.19\columnwidth]{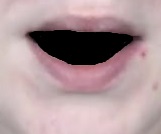}
        \\
        \makecell[l]{The corresponding \\ 3D head (ID: S16)}
        & \includegraphics[width=0.19\columnwidth]{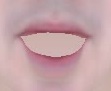}
        & \includegraphics[width=0.19\columnwidth]{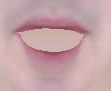}
        & \includegraphics[width=0.19\columnwidth]{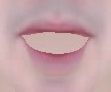}
        \\
        \makecell[l]{The non-corresponding \\ middle 3D head \\ (ID: S20)}
        & \includegraphics[width=0.19\columnwidth]{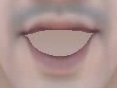}
        & \includegraphics[width=0.19\columnwidth]{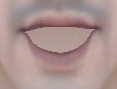}
        & \includegraphics[width=0.19\columnwidth]{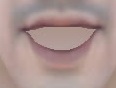}
     \end{tabular}
     }
    }\end{footnotesize}
    \caption{Consecutive frames of the phoneme /ih/ during utterance of the word "bin" from sentence "bin white with M 2 soon" for a real speaker (ID: S16) who classified under the low class of index 10 (mouth width), the corresponding 3D head (second row), the non-corresponding high 3D head (third row).} 
    \label{fig:Low_High_In10_S16}
\end{center}
\end{figure}

The RMSE results varied when 2D videos of the real speakers were mapped to the non-corresponding middle 3D heads and the non-corresponding high 3D heads. For mapping between 2D videos of real speakers and the non-corresponding middle 3D heads, the 3D head of a real speaker (ID: S32) gave the lowest score for both width and height because its middle mouth height was similar to or the same as most of the non-corresponding middle 3D heads with middle to high (IDs: S7, S15 and S24) or middle (ID: S55) mouth heights. What is striking in Table \ref{tab:low_In10} is that the corresponding 3D head of a real speaker (ID: S32) gave the lowest RMSE score for both width and height. Additionally, some of the non-corresponding high 3D heads gave the lowest scores for width for real speakers with similar mouth heights. For example, the non-corresponding high 3D head (ID: S20) gave the lowest score for a real speaker (ID: S16) with the same mouth height (i.e. low to middle). Figure \ref{fig:Low_High_In10_S16} gives an example of consecutive frames for the real speaker (ID: S16), the corresponding 3D head and the non-corresponding high 3D head (ID: S20) during utterance of the phoneme /ih/ of the word ``bin" from the phrase ``bin white with M 2 soon". Also, the non-corresponding high 3D head (ID: S22) gave the lowest score for a real speaker (ID: S35); this may because they both had high mouth heights. However, a t-test suggested a significant difference in the RMSE results between the corresponding 3D head of a real speaker (ID: S32) and the non-corresponding high 3D heads for both width and height. There was also a significant difference between the corresponding 3D head of a real speaker (ID: S38) and the non- corresponding high 3D heads for width.

\begin{figure}[htb]
\begin{center}
    \begin{footnotesize}{
    \resizebox{0.7\linewidth}{!}{
    \begin{tabular}{m{0.30\columnwidth}m{0.17\columnwidth}m{0.17\columnwidth}}
        \makecell[l]{Real speaker \\(ID: S38)}
        & \includegraphics[width=0.19\columnwidth]{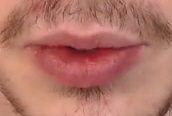}
        & \includegraphics[width=0.19\columnwidth]{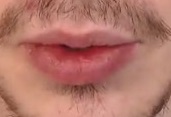}
        \\
        \makecell[l]{The corresponding \\ 3D head (ID: S38)}
        & \includegraphics[width=0.19\columnwidth]{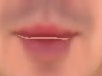}
        & \includegraphics[width=0.19\columnwidth]{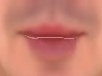}
        \\
        \makecell[l]{The non-corresponding \\ high 3D head \\ (ID: S54)}
        & \includegraphics[width=0.19\columnwidth]{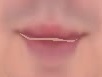}
        & \includegraphics[width=0.19\columnwidth]{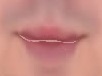}
        \\
        \makecell[l]{The non-corresponding \\ middle 3D head \\ (ID: S55)}
        & \includegraphics[width=0.19\columnwidth]{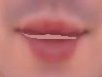}
        & \includegraphics[width=0.19\columnwidth]{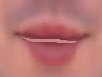}
         \\
        
        \makecell[l]{The non-corresponding \\ high 3D head \\ (ID: S22)}
        & \includegraphics[width=0.19\columnwidth]{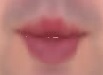}
        & \includegraphics[width=0.19\columnwidth]{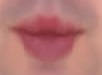}
        \end{tabular}
        }
    }\end{footnotesize}
    \caption{Consecutive frames of the phoneme /uw/ during utterance of the word ``two" from sentence ``bin white in I 2 soon" for a real speaker (ID: S38) who classified under the low class of index 10 (mouth width), the corresponding 3D head (second row), the non-corresponding middle 3D head (third row) and the non-corresponding high 3D heads (last two rows).} 
    \label{fig:Low_Mid_High_In10_S38}
\end{center}
\end{figure}

These finding confirm that 2D videos of real speakers who have middle or wide mouth widths can be used to animate 3D heads that associate with real speakers who have narrow mouth widths, as long as they have similar mouth heights, lip thicknesses and distances between the nose and the upper lip. For example, Figure \ref{fig:Low_Mid_High_In10_S38} gives an example of consecutive frames of a real speaker (ID: S38) classified in the low class of index 10, the corresponding 3D heads and the non-corresponding 3D heads. This Figure reveals that the non-corresponding high 3D head (ID: S54) produced a mouth shape more similar to the real speaker than the non-corresponding middle 3D head (ID: S55) due to closer classes of indices 7, 8, 9 and 12 of the corresponding real speakers, while the non-corresponding high 3D head (ID: S22) failed to give a more accurate shape because of its lip thickness.

\begin{table*}[htb]
\resizebox{\linewidth}{!}{
\begin{tabular}{cccccccccccllcc}
\hline
 & \multicolumn{2}{c}{} & \multicolumn{10}{c}{\begin{tabular}[c]{@{}c@{}}Non-corresponding \\ 3D head (middle)\end{tabular}} & \multicolumn{2}{c}{} \\ \cline{4-13}
 & \multicolumn{2}{c}{{\begin{tabular}[c]{@{}c@{}}Corresponding 3D \\ head (middle)\end{tabular}}} & \multicolumn{2}{c}{S7} & \multicolumn{2}{c}{S15} & \multicolumn{2}{c}{S24} & \multicolumn{2}{c}{S48} & \multicolumn{2}{c}{S55} & \multicolumn{2}{c}{{\begin{tabular}[c]{@{}c@{}}T-test\\ (P value)\end{tabular}}} \\ \cline{2-15} 
{2D video} & W & H & W & H & W & H & W & H & W & H & W & H & W & H \\ \hline
S7 &\textbf{ 0.273} & 0.128 & \cellcolor[HTML]{C0C0C0}{\color[HTML]{C0C0C0} } & \cellcolor[HTML]{C0C0C0} & 0.318 & 0.140 & 0.279 & \textbf{0.124} & 0.296 & 0.185 & 0.302 & 0.148 & 0.6312 & 0.1985 \\ \hline
S15 & \textbf{0.131} & \textbf{0.087} &0.158 & 0.191 & \cellcolor[HTML]{C0C0C0} & \cellcolor[HTML]{C0C0C0} & 0.186 & 0.132 & 0.164 & 0.151 & 0.179 & 0.134 & 0.0081 & 0.0177 \\ \hline
S24 & 0.219 & 0.123 & \textbf{0.172} & 0.142 & 0.237 & 0.122 & \cellcolor[HTML]{C0C0C0} & \cellcolor[HTML]{C0C0C0} & 0.245  & 0.142 & 0.260 & \textbf{0.118} & 0.6584 & 0.3001 \\ \hline
S48 & 0.149 & \textbf{0.071} & 0.158 & 0.188 & \textbf{0.133} & 0.081 & 0.144 & 0.095 & \cellcolor[HTML]{C0C0C0} & \cellcolor[HTML]{C0C0C0} & 0.166 & 0.072 & 0.8754 & 0.2506 \\ \hline
S55 & 0.143 & 0.167 & 0.118 & 0.116 & 0.119 & \textbf{0.103} & 0.122 & 0.104 & \textbf{0.113} &0.107& \cellcolor[HTML]{C0C0C0} & \cellcolor[HTML]{C0C0C0} & 0.1241 & 0.3809  \\ \hline
\multicolumn{1}{l}{} & \multicolumn{1}{l}{} & \multicolumn{1}{l}{} & \multicolumn{10}{c}{\begin{tabular}[c]{@{}c@{}}Non-corresponding \\ 3D head (low)\end{tabular}} & \multicolumn{1}{l}{} & \multicolumn{1}{l}{} \\ \cline{4-13}
\multicolumn{1}{l}{} & \multicolumn{1}{l}{} & \multicolumn{1}{l}{} & \multicolumn{2}{c}{S16} & \multicolumn{2}{c}{S32} & \multicolumn{2}{c}{S35} & \multicolumn{2}{c}{S38} & \multicolumn{2}{c}{\cellcolor[HTML]{C0C0C0}} & \multicolumn{1}{l}{} & \multicolumn{1}{l}{} \\ \hline
\multicolumn{1}{l}{} & W & H & W & H & W & H & W & H & W & H & \cellcolor[HTML]{C0C0C0}& \cellcolor[HTML]{C0C0C0}& W & H \\ \hline
S7 & \textbf{0.273} & \textbf{0.128} & 0.310 & 0.138 & 0.295 & 0.149 & 0.298 & 0.131 & 0.293 & 0.129 & \cellcolor[HTML]{C0C0C0} & \cellcolor[HTML]{C0C0C0} & 0.0064 & 0.1481 \\ \hline
S15 & \textbf{0.131} & \textbf{0.087} & 0.179 & 0.139 & 0.183 & 0.130 & 0.175 & 0.139 & 0.174 & 0.136 & \cellcolor[HTML]{C0C0C0} & \cellcolor[HTML]{C0C0C0} & 0.0002 & 0.0002 \\ \hline
S24 & 0.219 & 0.123 &\textbf{0.217} & 0.116 & 0.255 & 0.124 & 0.275 & 0.117 & 0.228 & \textbf{0.112} & \cellcolor[HTML]{C0C0C0} & \cellcolor[HTML]{C0C0C0} & 0.1558 & 0.1046 \\ \hline
S48 & 0.149 & \textbf{0.071} & 0.122 & 0.090 & 0.157 & \textbf{0.071} & 0.157 & 0.101 & \textbf{0.113}& 0.127 & \cellcolor[HTML]{C0C0C0} & \cellcolor[HTML]{C0C0C0} & 0.4464 & 0.1105 \\ \hline
S55 &\textbf{ 0.113} & 0.107 & 0.145 & 0.109 & 0.154 & \textbf{0.105} & 0.116 & 0.114 & 0.120& 0.116 & \cellcolor[HTML]{C0C0C0} & \cellcolor[HTML]{C0C0C0} & 0.1122 & 0.2056 \\ \hline
\multicolumn{1}{l}{} & \multicolumn{1}{l}{} & \multicolumn{1}{l}{} & \multicolumn{10}{c}{\begin{tabular}[c]{@{}c@{}}Non-corresponding\\ 3D head (high)\end{tabular}} & \multicolumn{1}{l}{} & \multicolumn{1}{l}{} \\ \cline{4-13}
\multicolumn{1}{l}{} & \multicolumn{1}{l}{} & \multicolumn{1}{l}{} & \multicolumn{2}{c}{S19} & \multicolumn{2}{c}{S20} & \multicolumn{2}{c}{S22} & \multicolumn{2}{c}{S46} & \multicolumn{2}{c}{S54} & \multicolumn{1}{l}{} & \multicolumn{1}{l}{} \\ \hline
 & W & H & W & H & W & H & W & H & W & H & \multicolumn{1}{c}{W} & \multicolumn{1}{c}{H} & W & H \\ \hline
S7 & 0.273 & 0.128 & \textbf{0.180} & 0.122 & 0.257 & 0.137 & 0.256 &\textbf{0.121} & 0.274 & \textbf{0.121} & \multicolumn{1}{c}{0.325} & \multicolumn{1}{c}{0.126} & 0.5645 & 0.4410 \\ \hline
S15 & \textbf{0.131} & \textbf{0.087} & 0.146 & 0.136 & 0.173 & 0.134 & 0.164 & 0.163 & 0.166 & 0.136 & \multicolumn{1}{c}{0.170} & \multicolumn{1}{c}{0.149} & 0.0022 & 0.0005 \\ \hline
S24 & \textbf{0.219} & 0.123 & 0.244 & 0.133 & 0.251 & 0.129 & 0.255 & 0.133 & 0.258 & 0.120 & \multicolumn{1}{c}{0.228} & \multicolumn{1}{c}{\textbf{0.114}} & 0.0062 & 0.5007 \\ \hline
S48 & 0.149 & 0.071 & 0.184 & 0.078 & 0.156 & \textbf{0.068} & 0.164 & 0.145 & 0.156 & 0.096 & \multicolumn{1}{c}{\textbf{0.139}} & \multicolumn{1}{c}{0.104} & 0.2129 & 0.1107 \\ \hline
S55 & 0.113 & 0.107 & \textbf{0.097} & 0.111 & 0.114 & \textbf{0.099} & 0.099 & 0.127 & 0.116 & 0.110 & \multicolumn{1}{c}{0.134} & \multicolumn{1}{c}{0.117} & 0.8886 & 0.2747 \\ \hline
\end{tabular}}
\caption{The RMS error averaged over 4 sentences for width (W) and height (H) of the mouth of the real speakers classified under the middle class of index 10 (mouth width), their corresponding 3D heads and the non-corresponding 3D heads. Values in bold means the lowest RMS error. The last column shows p value of the t-test results between each corresponding 3D head and the non-corresponding 3D heads for width and height.}
\label{tab:middle_In10}
\end{table*}

\begin{figure}[htb]
\begin{center}
    \begin{footnotesize}{
    \resizebox{0.7\linewidth}{!}{
    \begin{tabular}{m{0.30\columnwidth}m{0.17\columnwidth}m{0.17\columnwidth}m{0.17\columnwidth}}
        \makecell[l]{Real speaker \\(ID: S24)}
        & \includegraphics[width=0.19\columnwidth]{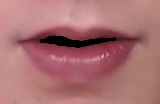}
        & \includegraphics[width=0.19\columnwidth]{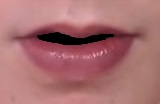}
        & \includegraphics[width=0.19\columnwidth]{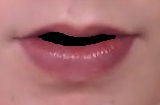}
        \\
        \makecell[l]{The corresponding \\ 3D head (ID: S24)}
        & \includegraphics[width=0.19\columnwidth]{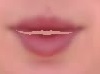}
        & \includegraphics[width=0.19\columnwidth]{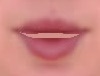}
        & \includegraphics[width=0.19\columnwidth]{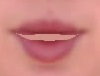}
        \\
        \makecell[l]{The non-corresponding \\ low 3D head \\ (ID: S32)}
        & \includegraphics[width=0.19\columnwidth]{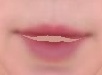}
        & \includegraphics[width=0.19\columnwidth]{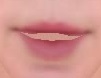}
        & \includegraphics[width=0.19\columnwidth]{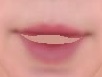}
        \\
        \makecell[l]{The non-corresponding \\ low 3D head \\ (ID: S35)}
        & \includegraphics[width=0.19\columnwidth]{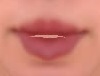}
        & \includegraphics[width=0.19\columnwidth]{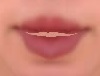}
        & \includegraphics[width=0.19\columnwidth]{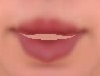}
         \\
        
        \makecell[l]{The non-corresponding \\ high 3D head \\ (ID: S19)}
        & \includegraphics[width=0.19\columnwidth]{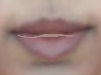}
        & \includegraphics[width=0.19\columnwidth]{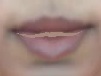}
        & \includegraphics[width=0.19\columnwidth]{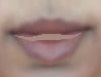}
        \end{tabular}
    }
    }\end{footnotesize}
    \caption{Consecutive frames of the phoneme /th/ during utterance of the word ``three" from sentence ``bin white in N 3 now" for a real speaker (ID: S24) who classified under the middle class of index 10 (mouth width), the corresponding 3D head (second row), the non-corresponding low 3D head (third and forth rows) and the non-corresponding high 3D heads (last row).} 
    \label{fig:Mid_Low_High_In10_S24}
\end{center}
\end{figure}

Table \ref{tab:middle_In10} shows the RMSE results averaged over four sentences for the width and height of the mouth aperture of real speakers classified in the middle class of index 10, their corresponding 3D heads, non-corresponding low, non-corresponding middle and non-corresponding high 3D heads. From this Table, it can be observed that three of the corresponding heads (IDs: S7, S15 and S55) gave the lowest scores for width, when the 2D videos of real speakers were mapped to the non-corresponding low 3D heads and the non-corresponding middle 3D heads. However, t-test results showed a significant difference in RMSE results between the corresponding 3D head of a real speaker (ID: S15) versus all the non-corresponding 3D heads for height and width. Also, there was a significant difference in the RMSE results for width between the corresponding 3D head of a real speaker (ID: S7) and the non-corresponding low 3D heads and between the corresponding 3D head of a real speaker (ID: S24) and the non-corresponding high 3D heads. These findings may prove that 2D videos of real speakers who have middle mouth width can be used to animate 3D heads that correspond to real speakers that have narrow, middle or wide mouth widths, as long as they have similar lip thicknesses or distances between the nose and the upper lip. Figure \ref{fig:Mid_Low_High_In10_S24} shows an example of consecutive frames of the phoneme /th/ from the word ``three" during phrase ``bin white in N 3 now" by a real speaker (ID: S24) classified in the middle class of index 10, the corresponding 3D heads, the non-corresponding low 3D heads and the non-corresponding high 3D heads. This figure shows how the 3D heads gave the correct mouth shape regardless of the mouth width. The mouth aperture of the non-corresponding middle 3D head (ID: S32) (third row) is slightly wide compared to the real speaker due to its middle mouth height (index 7).

Table \ref{tab:high_In10} shows the RMSE results averaged over four sentences for the width and height of the mouth aperture of real speakers classified in the high class of index 10 (mouth width), their corresponding 3D heads, non-corresponding low, non-corresponding middle and non-corresponding high 3D heads. From this Table, it can be noticed that the corresponding 3D head (ID: S19) gave the lowest RMSE scores for width for most of the speakers when it was mapped to 2D videos of real speakers who are classified under the high class. This may be due to the middle distance between the nose tip and the upper lip and the high mouth height. This also explains the significant difference suggested by the t-test result for height. Also, it can be noticed that the non- corresponding low 3D head (ID: S16) gave the lowest RMSE score for width for most of the speakers. This is probably due to the large distance between the nose tip and the upper lip. Another notable finding shown in this Table is that the non-corresponding middle 3D head (ID: S 7) gave the lowest RMSE score for most of the speakers. This may be due to the middle to high mouth height (index 7), upper lip thickness (index 8), lower lip thickness (index 9) and distance between the nose tip and the upper lip (index 12). The t-test results suggested no significant difference for width and height between four out of five of the corresponding 3D heads, the non-corresponding high 3D heads and the non-corresponding middle 3D heads, while there is no significant difference between the non-corresponding low 3D heads, the corresponding 3D heads for height and three out of five of the corresponding 3D heads for width. 

\begin{table*}[htb]
\resizebox{\linewidth}{!}{
\begin{tabular}{cccccccccccllcc}
\hline
 & \multicolumn{2}{c}{} & \multicolumn{10}{c}{\begin{tabular}[c]{@{}c@{}}Non-corresponding \\ 3D head (high)\end{tabular}} & \multicolumn{2}{c}{} \\ \cline{4-13}
 & \multicolumn{2}{c}{{\begin{tabular}[c]{@{}c@{}}Corresponding 3D \\ head (high)\end{tabular}}} & \multicolumn{2}{c}{S19} & \multicolumn{2}{c}{S20} & \multicolumn{2}{c}{S22} & \multicolumn{2}{c}{S46} & \multicolumn{2}{c}{S54} & \multicolumn{2}{c}{{\begin{tabular}[c]{@{}c@{}}T-test\\ (P value)\end{tabular}}} \\ \cline{2-15} 
{2D video} & W & H & W & H & W & H & W & H & W & H & W & H & W & H \\ \hline
S19 &\textbf{ 0.215} & \textbf{0.122} & \cellcolor[HTML]{C0C0C0}{\color[HTML]{C0C0C0} } & \cellcolor[HTML]{C0C0C0} & 0.218 & 0.129 & 0.222 & 0.134 & 0.247 & 0.128 & 0.238 & 0.140 & 0.0967 & 0.0297 \\ \hline
S20 & \textbf{0.239} & 0.166 &0.240 & \textbf{0.149} & \cellcolor[HTML]{C0C0C0} & \cellcolor[HTML]{C0C0C0} & 0.275 & 0.208 & 0.333 & 0.162 & 0.252 & 0.152 & 0.1933 & 0.9064 \\ \hline
S22 & 0.172 & 0.137 & 0.125 & 0.184 & \textbf{0.117 }& \textbf{0.134} & \cellcolor[HTML]{C0C0C0} & \cellcolor[HTML]{C0C0C0} & 0.149  & 0.156 & 0.147 & 0.156 & 0.0182 & 0.1391 \\ \hline
S46 & 0.231 & 0.118 & \textbf{0.166} & 0.136 & 0.216 &0.172 & 0.227 &\textbf{0.070} & \cellcolor[HTML]{C0C0C0} & \cellcolor[HTML]{C0C0C0} & 0.255 & 0.104 & 0.4785 & 0.9160 \\ \hline
S54 & 0.281 & 0.132 & \textbf{0.209} & 0.147 & 0.242 & 0.166 & 0.240 & \textbf{0.129} & 0.250 &0.131& \cellcolor[HTML]{C0C0C0} & \cellcolor[HTML]{C0C0C0} & 0.0148 & 0.2814  \\ \hline
\multicolumn{1}{l}{} & \multicolumn{1}{l}{} & \multicolumn{1}{l}{} & \multicolumn{10}{c}{\begin{tabular}[c]{@{}c@{}}Non-corresponding \\ 3D head (low)\end{tabular}} & \multicolumn{1}{l}{} & \multicolumn{1}{l}{} \\ \cline{4-13}
\multicolumn{1}{l}{} & \multicolumn{1}{l}{} & \multicolumn{1}{l}{} & \multicolumn{2}{c}{S16} & \multicolumn{2}{c}{S32} & \multicolumn{2}{c}{S35} & \multicolumn{2}{c}{S38} & \multicolumn{2}{c}{\cellcolor[HTML]{C0C0C0}} & \multicolumn{1}{l}{} & \multicolumn{1}{l}{} \\ \hline
\multicolumn{1}{l}{} & W & H & W & H & W & H & W & H & W & H & \cellcolor[HTML]{C0C0C0}& \cellcolor[HTML]{C0C0C0}& W & H \\ \hline
S19 & \textbf{0.215} & 0.122 & 0.268 & 0.130 & 0.262 & 0.125 & 0.250 & \textbf{0.120} & 0.242 & 0.151 & \cellcolor[HTML]{C0C0C0} & \cellcolor[HTML]{C0C0C0} & 0.0062 & 0.2575 \\ \hline
S20 & 0.239 & 0.166 & \textbf{0.223} & \textbf{0.155} & 0.326 & 0.157 & 0.327 & 0.173 & 0.281 & 0.156 & \cellcolor[HTML]{C0C0C0} & \cellcolor[HTML]{C0C0C0} & 0.1332 & 0.2708 \\ \hline
S22 & 0.172 & \textbf{0.137} &\textbf{0.124} & 0.138 & 0.300 & 0.233 & 0.127 & 0.157 & 0.147 & 0.142 & \cellcolor[HTML]{C0C0C0} & \cellcolor[HTML]{C0C0C0} & 0.9564 & 0.2634 \\ \hline
S46 & \textbf{0.231} & 0.118 & 0.256 & 0.105 & 0.269 & 0.191 & 0.244 & 0.108 & 0.248& \textbf{0.099} & \cellcolor[HTML]{C0C0C0} & \cellcolor[HTML]{C0C0C0} & 0.0244 & 0.7461 \\ \hline
S54 &0.281 & \textbf{0.132} & \textbf{0.255} & 0.135 & \textbf{0.255} & 0.152 & 0.267 & 0.136 & 0.293 & 0.133 & \cellcolor[HTML]{C0C0C0} & \cellcolor[HTML]{C0C0C0} & 0.2289 & 0.2081 \\ \hline
\multicolumn{1}{l}{} & \multicolumn{1}{l}{} & \multicolumn{1}{l}{} & \multicolumn{10}{c}{\begin{tabular}[c]{@{}c@{}}Non-corresponding\\ 3D head (middle)\end{tabular}} & \multicolumn{1}{l}{} & \multicolumn{1}{l}{} \\ \cline{4-13}
\multicolumn{1}{l}{} & \multicolumn{1}{l}{} & \multicolumn{1}{l}{} & \multicolumn{2}{c}{S7} & \multicolumn{2}{c}{S15} & \multicolumn{2}{c}{S24} & \multicolumn{2}{c}{S48} & \multicolumn{2}{c}{S55} & \multicolumn{1}{l}{} & \multicolumn{1}{l}{} \\ \hline
 & W & H & W & H & W & H & W & H & W & H & \multicolumn{1}{c}{W} & \multicolumn{1}{c}{H} & W & H \\ \hline
S19 &0.215 & \textbf{0.122} & \textbf{0.209} & 0.169 & 0.244 & 0.130 & 0.254 &\textbf{0.122} & 0.214 & 0.130 & \multicolumn{1}{c}{0.281} & \multicolumn{1}{c}{0.131} & 0.1284 & 0.1582 \\ \hline
S20 & 0.239 & 0.166 & \textbf{0.208} & 0.350 & 0.274 & 0.158 & 0.308 & 0.171 & 0.294 & \textbf{0.148} & \multicolumn{1}{c}{0.309} & \multicolumn{1}{c}{0.163} & 0.1023 & 0.4491 \\ \hline
S22 & 0.172 & 0.137 & 0.140 & \textbf{0.116} & \textbf{0.110} & 0.135 & 0.129 & 0.146 & 0.179 & 0.153 & \multicolumn{1}{c}{0.168} & \multicolumn{1}{c}{\textbf{0.157}} & 0.1011 & 0.5825 \\ \hline
S46 &0.231 & 0.118 & \textbf{0.203} & \textbf{0.116} & 0.255 & 0.149 & 0.261 & 0.122 & 0.215 & 0.138 & \multicolumn{1}{c}{0.252} & \multicolumn{1}{c}{0.144} & 0.6259 & 0.0679 \\ \hline
S54 & 0.281 & \textbf{0.132} & \textbf{0.240} & 0.156 & 0.273 & 0.155 & 0.254 & 0.142 & 0.273 & 0.133 & \multicolumn{1}{c}{0.256} & \multicolumn{1}{c}{0.149} & 0.0255 & 0.0252 \\ \hline
\end{tabular}}
\caption{The RMS error averaged over 4 sentences for width (W) and height (H) of the mouth of the real speakers classified under the high class of index 10 (mouth width), their corresponding 3D heads and the non-corresponding 3D heads. Values in bold means the lowest RMS error. The last column shows p value of the t-test results between each corresponding 3D head and the non-corresponding 3D heads for width and height.}
\label{tab:high_In10}
\end{table*}

\begin{figure}[htb]
\begin{center}
    \begin{footnotesize}{
    \resizebox{0.7\linewidth}{!}{
    \begin{tabular}{m{0.30\columnwidth}m{0.17\columnwidth}m{0.17\columnwidth}}
        \makecell[l]{Real speaker\\ (ID: S54)}
        & \includegraphics[width=0.19\columnwidth]{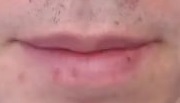}
        & \includegraphics[width=0.19\columnwidth]{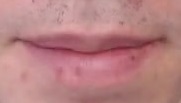}
        \\
        \makecell[l]{The corresponding \\ 3D head (ID: S54)}
        & \includegraphics[width=0.19\columnwidth]{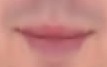}
        & \includegraphics[width=0.19\columnwidth]{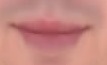}
        \\
        \makecell[l]{The non-corresponding \\ low 3D head \\ (ID: S38)}
        & \includegraphics[width=0.19\columnwidth]{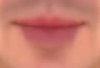}
        & \includegraphics[width=0.19\columnwidth]{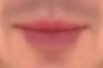}
        \\
        \makecell[l]{The non-corresponding \\ middle 3D head \\ (ID: S55)}
        & \includegraphics[width=0.19\columnwidth]{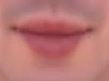}
        & \includegraphics[width=0.19\columnwidth]{figs/Mid_s54_s55_p_lwbb8a_18_C.jpg}
        \end{tabular}
    }
    }\end{footnotesize}
    \caption{Consecutive frames of the phoneme /b/ during utterance of the letter "B" from sentence "lay white by B 8 again" for a real speaker (ID: S54) who classified under the high class of index 10 (mouth width), the corresponding 3D head (second row), the non-corresponding low 3D head (third row) and the non-corresponding middle 3D heads (last row).} 
    \label{fig:High_Mid_Low_In10_S54}
\end{center}
\end{figure}

These findings suggest that animating the wide mouths of 3D heads can be achieved using 2D videos of real speakers who have narrow or middle mouth widths as long as they have similar lip thicknesses, shapes and distances between the nose and the upper lip. Figure \ref{fig:High_Mid_Low_In10_S54} shows an example of consecutive frames of a real speaker (ID: S54) classified in the high class of index 10, the corresponding 3D head, the non-corresponding low 3D head and the non-corresponding middle 3D head during utterance of the letter ``B" from the phrase ``lay white by B 8 again". This figure shows how all the 3D heads gave the correct mouth shape for the phoneme /b/, including the non-corresponding low 3D head.

\section{Subjective evaluation}
\label{subsec:Subjective_Evaluation_Ex6}

The subjective evaluation compares the naturalness of animations generated using the mapping process described in Section  \ref{subsec:Objective_Evaluation_Ex6}. This test investigates the impact of mouth height (index 7) and width (index 10) variation between real speakers and 3DMMs on the resulting 3D lip motions. 2D video of a real speaker from the Audio-Visual Lombard Grid corpus who is classified under one class of index 7 or index 10 are used to animate the corresponding 3D head and the non-corresponding 3D heads that relate to other speakers who are classified under the other two classes. Videos of the animated 3D heads are synchronised with the clean audio signal of the real speaker. The clean audio is used to enable the participants to judge the extent to which the animated lip movement is as smooth as a real speaker’s and how likely it was that the movement would produce those sounds. The evaluation addresses two main points:

\begin{itemize}
    \item The impact of differences between the real speakers and the 3D heads in mouth height and width on the resulting 3D lip animation.  
    \item The variation range in mouth height and width between real speakers and 3D heads that provides sufficiently-3D lip motion.
\end{itemize}

\subsection{Stimuli}
\begin{figure}[htb]
\begin{center}
{\includegraphics[trim =15mm 80mm 00mm 00mm, clip,width=3.1in,keepaspectratio]{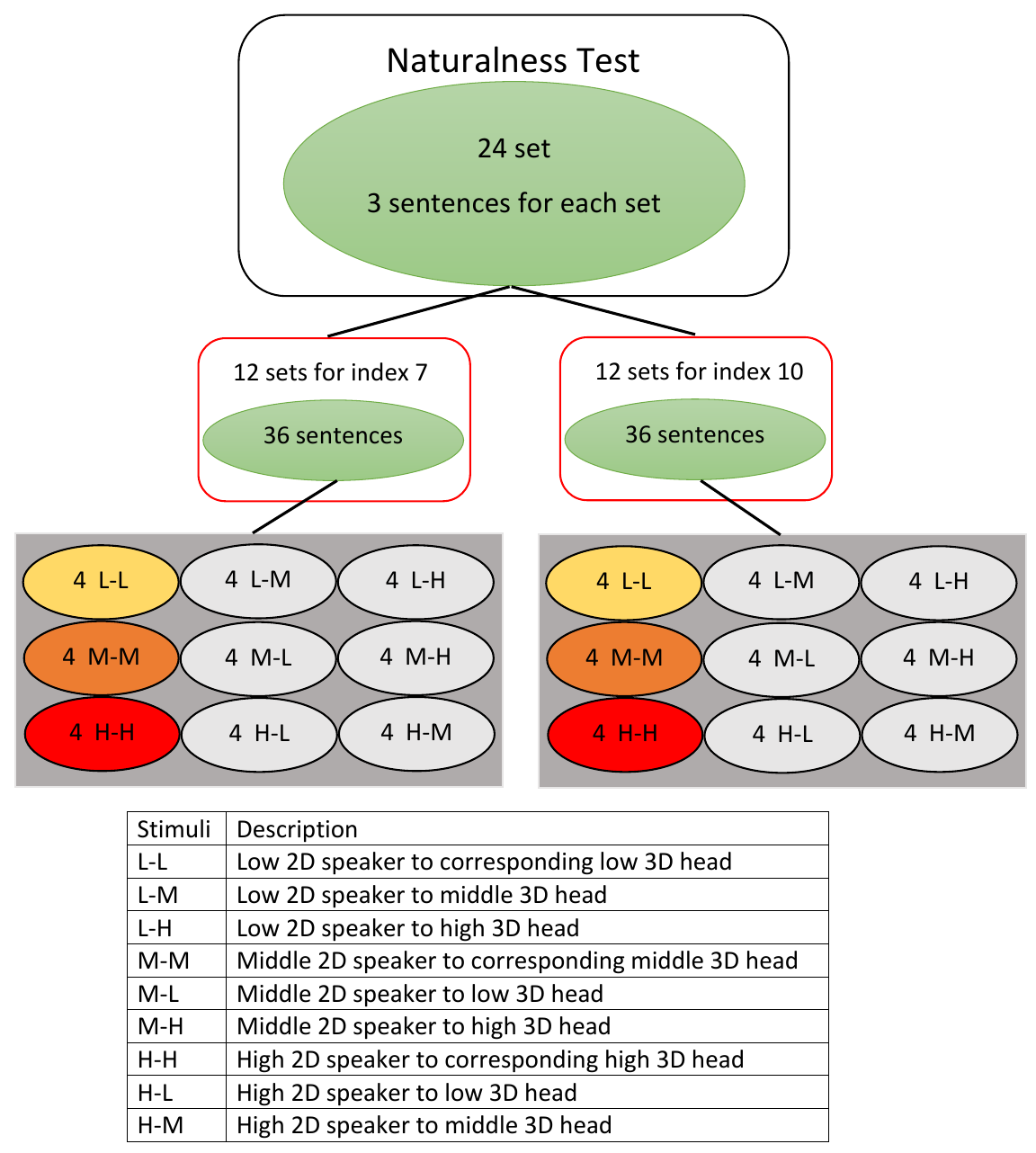}}{}
\caption{Structure of stimuli.}
\label{Fig:Stimuli_Structure}
\end{center}
\end{figure}

2D front view videos of 24 real speakers were selected to be mapped to the corresponding 3D heads and the non-corresponding 3D heads of the real speakers who are classified under the other two classes. Twelve speakers were selected based on classes of index 7, while the other twelve speakers were selected based on classes of index 10 (Four speakers from each class of each index as shown in Figure \ref{Fig:Stimuli_Structure}).     

Based on this, 72 2D videos of 3D talking heads were used in this test. Three separate animations were presented side by side for each set (24 sets in total presented in a random order for each participant). Twelve sets showed the resulting animation for mapping between non-corresponding faces based on the classes of index 7, while the other 12 sets showed the resulting animation for mapping between non-corresponding faces based on the classes of index 10. For each set, three 2D videos of 3D animations were presented side by side. All three heads were animated using 2D videos of one real speaker classified in one of the three classes. One animation corresponded to that real speaker, and the other two corresponded to different real speakers classified in the other two classes. Figure \ref{Fig:Stimuli_Structure} illustrates the structure of the stimuli. The participants used a play button to repeat each sentence and watched each video three times. After each set of videos, the participants were asked to choose which 3D talking head had the most natural lip movements and which had the least. The selection scores of a subject for the best and the worst choices were used to evaluate the impact of the mouth height and the mouth width on the resulting animation.

\subsection{Participants}

Two groups of participants with normal hearing and vision were recruited from the Department of Computer Science, University of Sheffield, and tested individually in an acoustically isolated booth with visual signals presented on a computer screen combined with acoustic signals presented binaurally through headphones. The first group consisted of 12 native English speakers $E$ and the second group consisted of 15 non-native English speakers $N$ (from different Arabic countries -- Bahrain, Egypt, Iraq, Libya and Saudi Arabia -- with  IELTS score $\epsilon$ [5.5,9]), where $E$ and $N$ denote native English speakers and non-native English speakers, respectively. It was deemed important to also test the results on non-native English speakers since such animation has been used for pronunciation training systems \cite{ali2015effects, dey2010talking, dey2010evaluation, engwall2012analysis, engwall2007pronunciation,  massaro1998perceiving, wang2012phoneme} and has been utilised more widely for customer services and entertainment such as films and games \cite{xie2016visual}. This study was ethically approved via the University of Sheffield’s ethics review procedure.

\subsection{Results}

\begin{figure}[htb]
\begin{center}
      \resizebox{1.0\linewidth}{!}{
        \includegraphics[trim = 02mm 00mm 120mm 01mm, clip,width=3.65in,keepaspectratio]{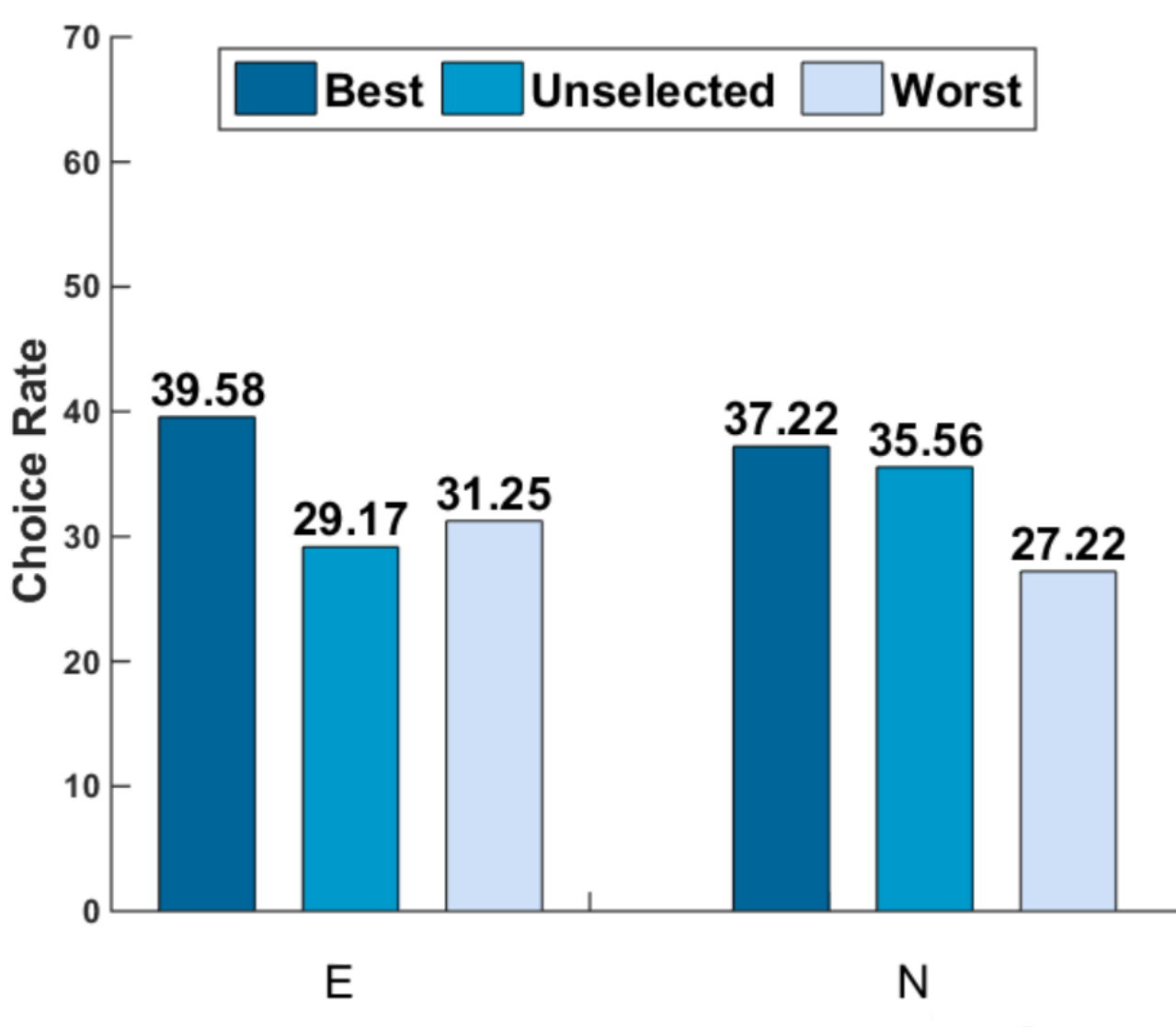}
        \includegraphics[trim = 02mm 00mm 120mm 00mm, clip,width=3.65in,keepaspectratio]{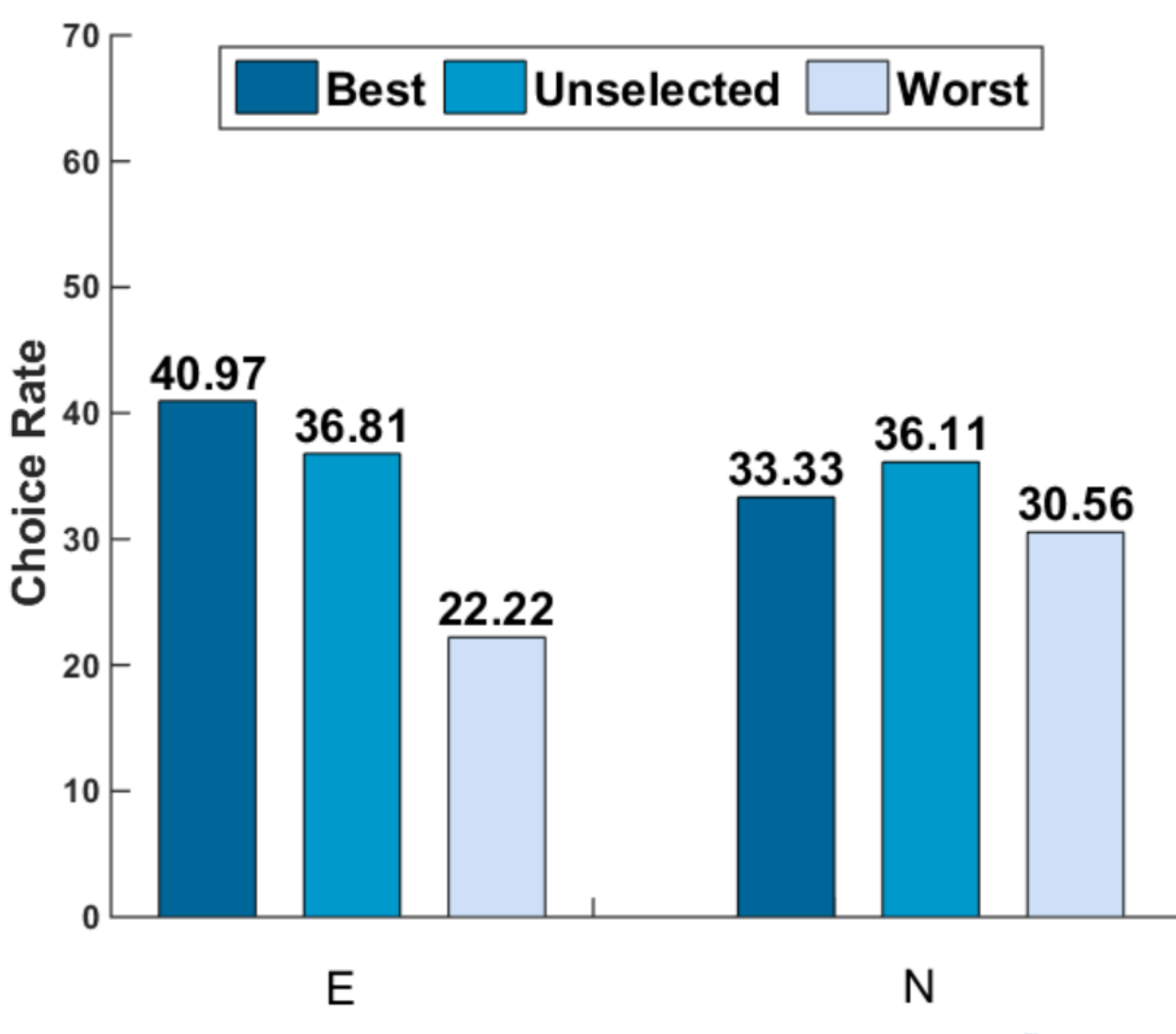}
      }
    \caption{Results for the best, not chosen and worst rates for the corresponding 3D heads for the $E$ and $N$ groups: Rates for index 7 (left); Rates for index 10 (right). }
    \label{Fig:Native_Arabic_Best_Moderate_Worst}
\end{center}
\end{figure}

Figure \ref{Fig:Native_Arabic_Best_Moderate_Worst} shows the most and least natural choices made by the two groups for the corresponding 3D heads for index 7 (left) and index 10 (right). Generally, the differences in mouth height (index 7) between real speakers and 3D head models have a greater influence on the resulting animation, where the two groups voted for the corresponding 3D heads as most natural with no significant difference as confirmed by a t-test result (p=0.6514). 

For the mouth width (index 10), the $E$ group outperformed the $N$ group for voting for the corresponding 3D heads as most natural. This may be because the stimuli were presented in their language, suggesting that the differences in mouth width have less impact on the 3D lip motions, which meant the non-native group ($N$) were not able to select the corresponding 3D heads as the best choice. However, no significant difference was found between the two groups for voting for the corresponding 3D heads as most natural (p=0.1482).

\begin{figure*}[htb]
\begin{center}
    \stackunder[3pt]{
      \resizebox{0.75\linewidth}{!}{%
        \includegraphics[trim = 00mm 06mm 06mm 00mm, clip,width=5.50in,keepaspectratio]{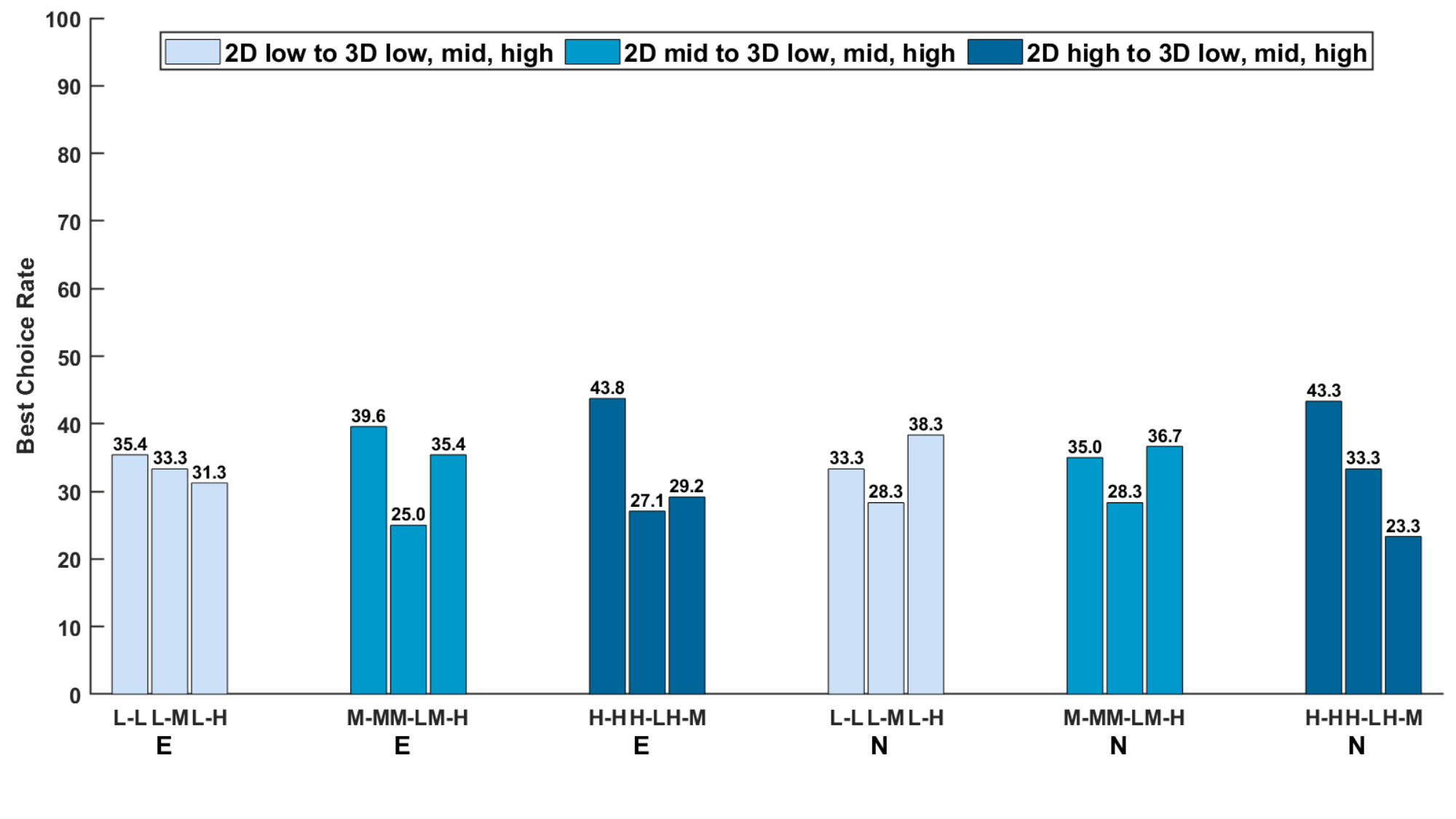}
      }
    }{}
    \stackunder[3pt]{
      \resizebox{0.75\linewidth}{!}{%
        \includegraphics[trim = 00mm 06mm 06mm 00mm, clip,width=5.50in,keepaspectratio]{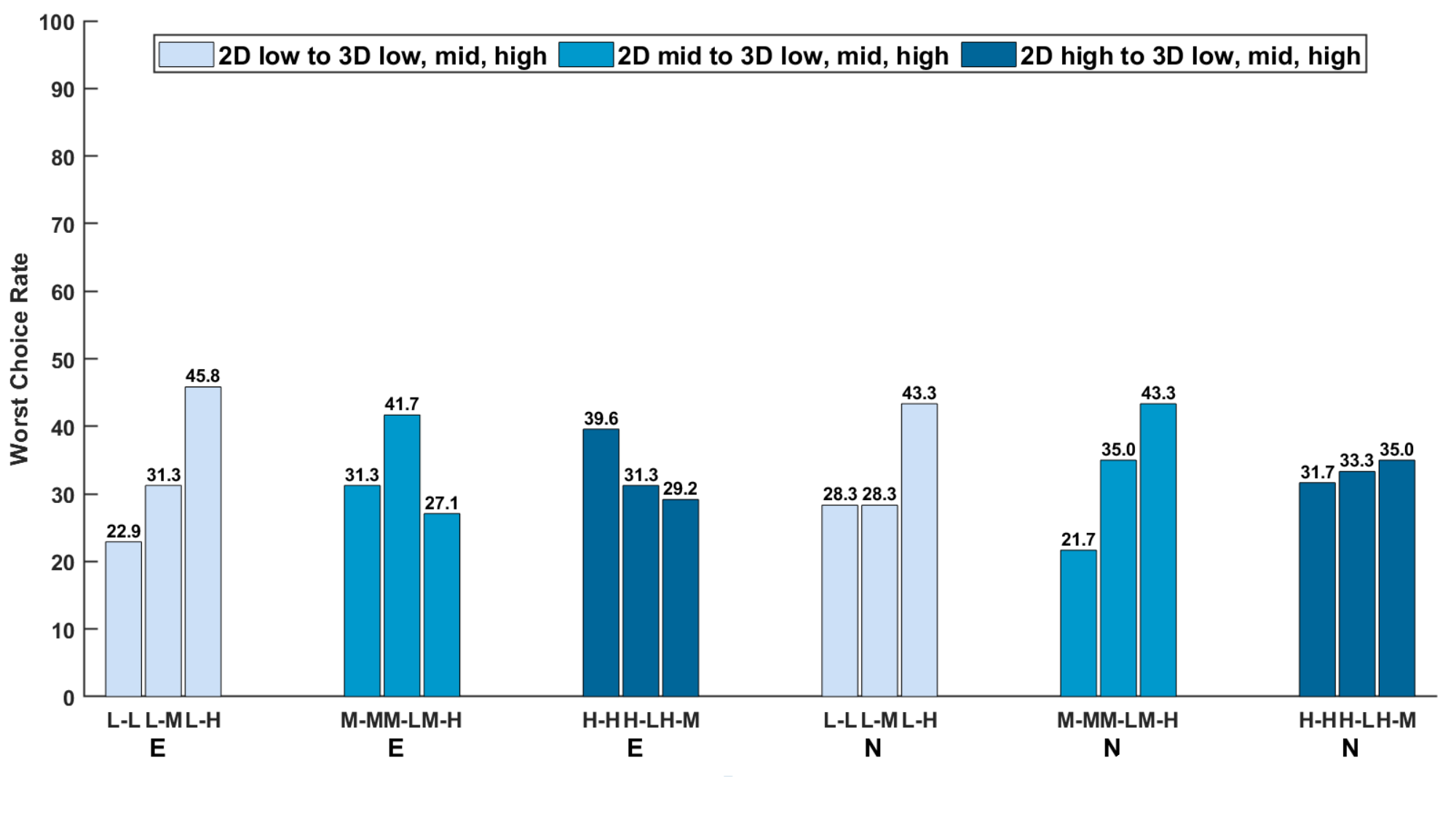}
      }
    }{}
    \caption{Results for the best and worst choice rates for the corresponding 3D heads and the non-corresponding 3D heads for each class of index 7 for the $E$ and $N$ groups: Rates of the best answer (top); Rates of the worst answer (bottom). }
    \label{Fig:Native_Arabic_Best_Worst_In7}
\end{center}
\end{figure*}

Figure \ref{Fig:Native_Arabic_Best_Worst_In7} shows the most and least natural choice rates for the corresponding 3D heads and the non-corresponding 3D heads for each class of index 7 for the two groups. The $E$ group found the corresponding 3D heads of each class to be most natural, while the $N$ voted only for the corresponding high 3D heads. The two groups were able to distinctively choose the corresponding high 3D heads as having the most natural lip motions. This is confirmed by a t-test result that showed no significant difference between the two groups for choosing the corresponding high 3D heads as having most natural lip motions (p=0.0981). However, for the high class, t-test results showed no significant difference between the corresponding 3D heads and the non-corresponding low 3D heads for the $E$ group (p=0.3225) and the $N$ group (p=0.1643). Also, no significant difference was found between the corresponding 3D heads and the non-corresponding middle 3D heads for the $E$ group (p=0.2229) and the $N$ group (p=0.0611). The two groups found the animation generated by mapping 2D videos of a real speaker classified in the low class to the non-corresponding high 3D heads to be the least natural. A t-test result showed no significant difference between the two groups for voting for the non-corresponding high 3D heads as least natural (p=0.7882). This confirms the objective test results provided in Table \ref{tab:Low_In7}. However, no significant difference was found between the non-corresponding high 3D heads and the corresponding low 3D heads for the two groups (p=0.1362 for $E$ and p=0.1442 for $N$) or between the non-corresponding high 3D heads and the corresponding middle 3D heads for the two groups (p=0.1891 for $E$ and p=0.1781 for $N$). 

\begin{figure*}[htb]
\begin{center}
    \stackunder[3pt]{
      \resizebox{0.75\linewidth}{!}{%
        \includegraphics[trim = 00mm 10mm 06mm 00mm, clip,width=5.50in,keepaspectratio]{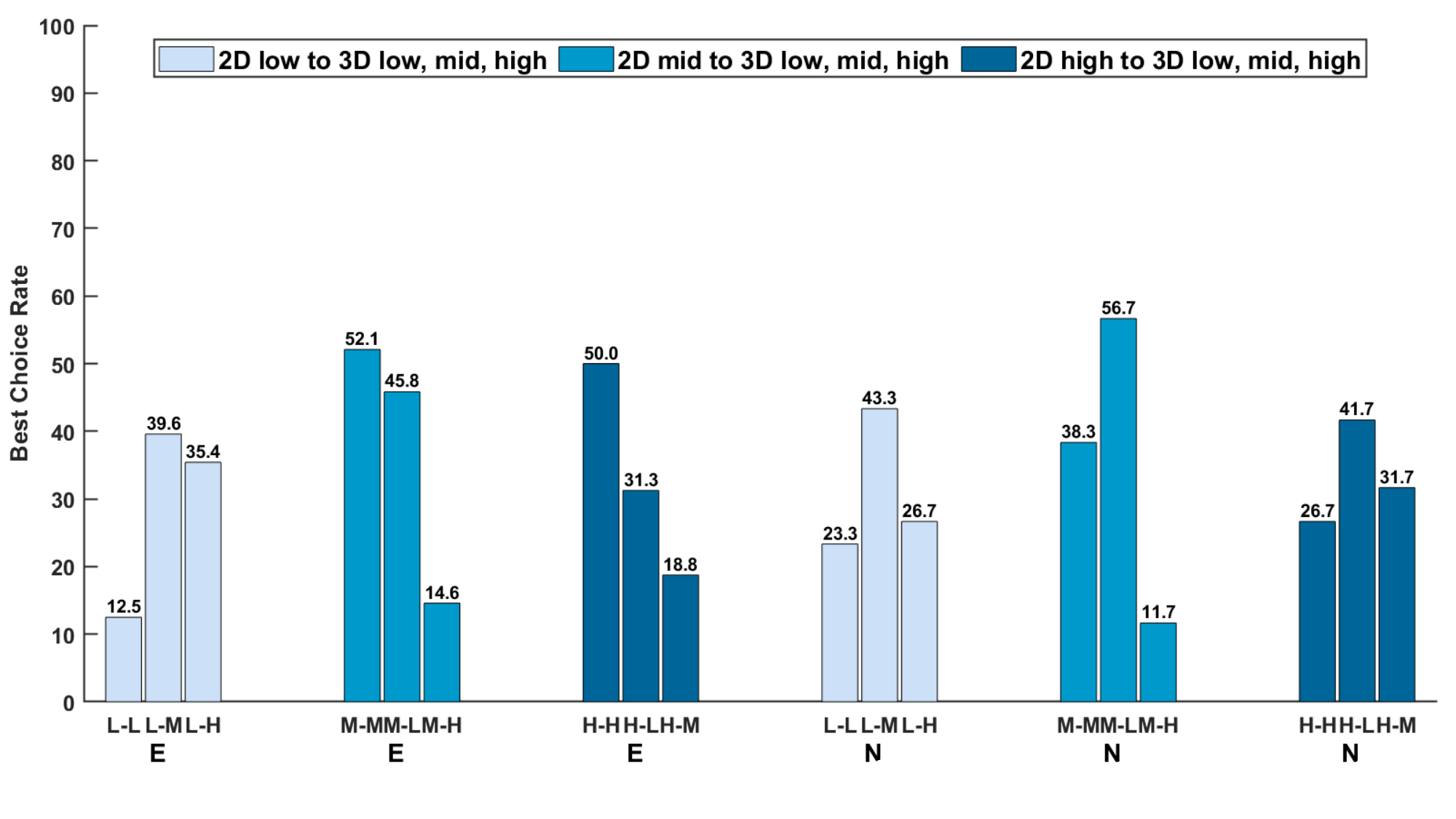}
      }
    }{}
    \stackunder[3pt]{
      \resizebox{0.75\linewidth}{!}{%
        \includegraphics[trim = 00mm 10mm 06mm 00mm, clip,width=5.50in,keepaspectratio]{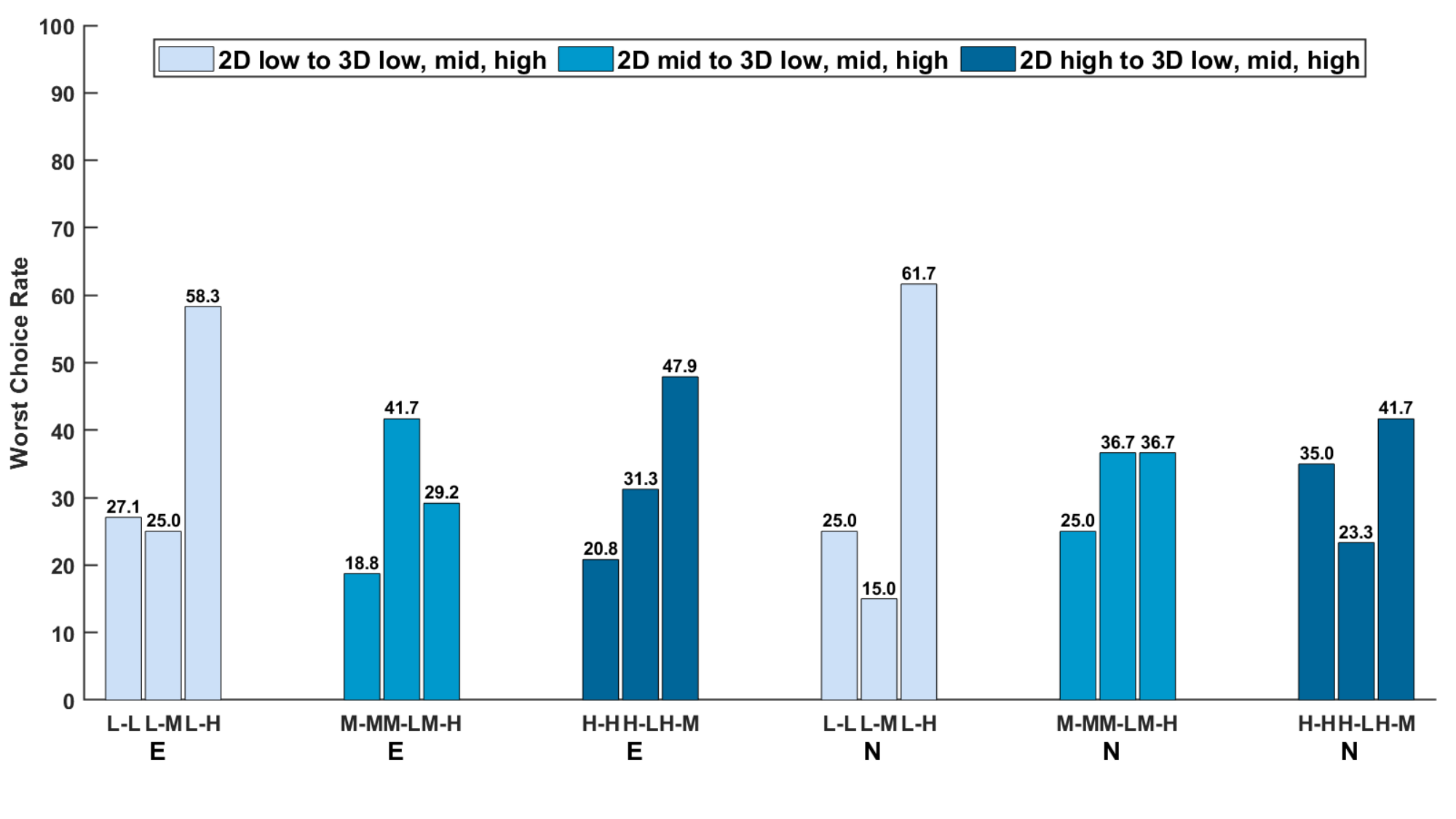}
      }
    }{}
    \caption{Results for the best and worst choice rates for the corresponding 3D heads and the non-corresponding 3D heads for each class of index 10 for the $E$ and $N$ groups: Rates of the best answer (top); Rates of the worst answer (bottom). }
    \label{Fig:Native_Arabic_Best_Worst_In10}
\end{center}
\end{figure*}

Figure \ref{Fig:Native_Arabic_Best_Worst_In10} shows the most and least natural choice rates for the corresponding 3D heads and the non-corresponding 3D heads for each class of index 10 for the two groups. For the high class, t-test results showed a significant difference between the corresponding 3D heads and the non-corresponding middle 3D heads for the $E$ group (p=0.0280). The $E$ group found the corresponding 3D heads of the middle and the high classes to be the most natural. For the low class, the two groups found the non-corresponding high 3D heads to have the least natural lip motions. This was confirmed by a t-test result that showed no significant difference between the two groups for selecting the non-corresponding high 3D head as having the least natural lip motions (p=0.7575). A significant difference was suggested by the t-test results between the corresponding 3D heads and the non-corresponding high 3D heads for the two groups (p=0.0448 for the $E$ group and p=0.0046 for the $N$ group). Also, a significant difference was found between the non-corresponding high 3D heads and the non-corresponding middle 3D heads for the two groups (p=0.0012 for the $E$ group and p=0.0009 for the $N$ group). This may prove that reasonable 3D lip motions cannot be achieved when 2D videos of real speakers with narrow mouth widths are mapped to 3D heads that relate to real speakers with wide mouth widths. 

\subsection{Discussion}

This study has investigated the impact of similarities and differences in the facial features between real speakers and 3DMMs on the resulting 3D lip motions and also defined the ranges of differences in facial features between real speakers and 3D heads that allow adequate 3D lip motions to be achieved.

It was found that native English-speaking participants were able to distinguish between the corresponding and non-corresponding 3D heads slightly better than non-native English-speaking participants for the two tested indices (indices 7, mouth height, and 10, mouth width). This may be because the native participants had greater linguistic competence than the non-natives. 

The two groups were able to distinguish between the corresponding 3D heads and the non-corresponding low 3D heads for the high class of index 7 (mouth height), where the participants chose the lip motions of the corresponding high 3D heads as the most natural. However, the results for the least natural choice for this class were contrary for the $N$ group and convergent for the $E$ group. This indicates that selecting one of three choices is more difficult than choosing between two options. This could also apply to the most natural choice answers for the low class, where the results were close for each 3D head alternate, although the participants were able to select the non-corresponding high 3D heads as the least natural. This indicates that the difference between the corresponding 3D low heads and the non-corresponding high 3D heads is distinguishable, which was also confirmed by the objective test results. 

For the low class of index 10 (mouth width), the two groups were able to distinctly choose the non-corresponding high 3D heads as the least natural, which confirms that the difference is significant and distinguishable between these classes for this index. These findings are not comparable with the objective test results due to an unbalanced number of male and female speakers in each class of the tested indices. Consequently, presenting all possible methods of mapping to the participants was restricted by this factor, as it is not reasonable to display a 3D head of a real male speaker combined with a female audio signal to the participants. This suggests that in order to accurately investigate the effects of differences and similarities in the facial features between real speakers and 3DMMs on the resulting 3D visual speech animation, a large amount of data is essential. However, the most natural choice answers for this class indicate confusion between the corresponding 3D heads and the non-corresponding 3D heads. The performance of the $N$ group for the high class of index 10 (mouth width) is mixed for the most and least natural choice answers; this may be because the variations in this index are not noticeable to non-native participants in comparison to index 7 (mouth height), which has a greater effect on lip closure.      

\section{Conclusions}
\label{sec:conclusion}

In this paper, an investigation of the effects of differences and similarities in facial features between real speakers and 3DMMs on the mapping process was presented. The facial features of real speakers were represented by 12 indices, and each index was classified into three classes: low, middle and high. In this paper, two indices representing mouth height (index 7) and width (index 10) were investigated separately by mapping between real speakers from different classes to their corresponding 3D heads and 3D heads that corresponded to different speakers in the same class or different classes.

The resulting 3D lip motions were evaluated quantitatively and qualitatively. The results of the quantitative test suggest that, for index 7 (mouth height), the mapping between real speakers with low mouth height and the 3D heads that correspond to real speakers with high mouth height, or vice versa, leads to unpleasant 3D lip motions. For index 10 (mouth width), the results varied between the classes, which confirms that mouth width does not have significant effect on the mapping process, whilst other facial features should be considered, such as lip thickness. The qualitative evaluation results suggest that native English-speaking participants are able to distinguish between the corresponding and non-corresponding 3D heads slightly better than non-native speakers. For the two tested indices, the two groups of participants chose the non-corresponding high 3D heads as having the most unnatural lip motions when they were mapped to real speakers classified in low classes. For index 7 (mouth height), the two groups selected the corresponding high 3D heads as having the most natural lip motions. This is not the case with index 10 (mouth width), where only the native-speaking participants were able to select the corresponding high 3D heads as having the most natural lip motions. This may confirm that mouth width does not have a considerable effect on the resulting 3D lip motions due to limited changes in this feature during speech in comparison with index 7 (mouth height), which affects lip closure.

Based on these findings, it is thus important that any mismatch between a real actor's mouth and the 3D synthetic character's mouth that is being animated should be considered carefully. This has implications for training systems for the hard of hearing and for animation production for entertainment applications. Future work will consider the impact of other aspects of the mouth and other facial proportions.

\section*{Declaration of competing interest}
The authors declare that they have no known competing financial interests or personal relationships that could have appeared to influence the work reported in this paper.

\section*{Acknowledgements}

This research has been supported by the Libyan Ministry of Higher Education and Derna University.

\bibliographystyle{ieeetr}
\bibliography{sample}

\end{document}